\documentclass[10pt, aps,pre,twocolumn,amsfonts,amssymb,amsmath,showkeys,floatfix]{revtex4-2}
\usepackage{graphicx}           % needed for figures
\usepackage[colorlinks,linkcolor=blue,citecolor=blue]{hyperref}
\usepackage[caption=false,subrefformat=parens,labelformat=parens]{subfig}
\usepackage{bm}          % for bold math
\usepackage{booktabs}
\usepackage{tabularx}
\usepackage{url}                % for on-line citations
\usepackage{color}
\usepackage{mathtools}
\usepackage{accsupp}            % Additional data for text narrator. e.g., <, =, and other symbols
\usepackage{listings}  
\usepackage{algorithm}
\usepackage{algpseudocode}

\renewcommand*{\vec}[1]{\bm{\mathrm{#1}}}  %bold vector\\
\newcommand*{\vechat}[1]{\vec{\hat{#1}}}  %bold unit vector

\usepackage{tikz}
\usetikzlibrary{patterns}
\newcommand{\hatched}{% Don't remove the percent symbols at the end of following lines!
\definecolor{red(pigment)}{rgb}{0.93, 0.11, 0.14}%
\textcolor{red(pigment)}{\hspace{1pt}%
\begin{tikzpicture}
\draw[pattern=north east lines, pattern color=red(pigment)] (0,0) rectangle (6pt,6pt);
\end{tikzpicture}}\hspace{1pt}}

\newcommand{\shaded}{\textcolor[rgb]{0.52,0.81,0.2}{$\hspace{0.5pt}\blacksquare\hspace{0.5pt}$}}
\newcommand{\subfigref}[2]{\hyperref[#1]{\ref*{#1}(#2)}}
\newcommand{\spacedsubfigref}[2]{\hyperref[#1]{\ref*{#1}~(#2)}}
\begin{document}

\title{Comparison of the clock, stochastic cutoff, and Tomita Monte Carlo methods in simulating the dipolar triangular lattice at criticality}

\author{S. Ismailzadeh}
\author{M. D. Niry}
\thanks{Author to whom correspondence should be addressed.
	Electronic address: \url{m.d.niry@iasbs.ac.ir}; 
	URL: \url{http://www.iasbs.ac.ir/~m.d.niry/}}
\affiliation{Department of Physics, Institute for Advanced Studies in Basic Sciences (IASBS), Zanjan 45137-66731, Iran}

\date{\today}

\begin{abstract}
Magnetic nanostructures find application in diverse technological domains and their behavior is significantly influenced by long-range dipolar interactions. However, simulating these systems using the traditional Metropolis Monte Carlo method poses high computational demand. Several methods, including the clock, stochastic cutoff, and Tomita approaches, can reduce the computational burden of simulating 2D systems with dipolar interactions. Although these three methods rely on distinct theoretical concepts, they all achieve complexity reduction by a common strategy. Instead of calculating the energy difference between a spin and all its neighbors, they evaluate the energy difference with only a limited number of randomly chosen neighbors. This is achieved through methods like the dynamic thinning and Fukui-Todo techniques. In this article, we compared the performance of the clock, SCO, Tomita, and Metropolis methods near the critical point of the dipolar triangular lattice to identify the most suitable algorithm for this type of simulation. Our findings show that while these methods are less suitable for simulating this system in their untuned implementation, incorporating the boxing nearby neighbors method and overrelaxation moves makes them significantly more efficient and better suited than the Metropolis method with overrelaxation.
\end{abstract}
\maketitle

\section{\label{sec:int}Introduction}
Two of the four fundamental forces are long-range. The Coulomb interaction plays a vital role from the subatomic to the molecular level while gravitational interaction dominates on a cosmic scale, so long-range interactions are widespread in natural phenomena. 
At the level of Coulomb interaction, the signed nature of charges becomes further challenging when creating vectorized dipoles in the anisotropic long-range dipolar interaction. The authors of this paper are motivated to address this significant challenge because the dipolar interaction is crucial in determining the behavior of nanomagnetic arrays, magnetic thin films, and nanostructures \cite{debell2000, kechrakos2008}, which have diverse technological applications in spintronics \cite{karmakar2011}, data storage \cite{sun2004}, and biomedicine \cite{peixoto2020}.

Simulating systems near the critical point is intriguing as it can uncover interesting emergent behaviors that occur during a continuous phase transition \cite{newman1999, kardar2007}. One of the main approaches for simulating a system near the critical point is the Monte Carlo (MC) method \cite{newman1999}. The most commonly used MC method is the Metropolis algorithm. However, when applied to systems with long-range interactions, close to the critical point, this algorithm faces two challenges \cite{flores-sola2017,luijten1995}.

One challenge lies in the computational cost of simulating these types of systems. This is attributed to the need to consider a significant number of interactions. Consequently, as the system size increases, the time required for computation grows at a rate proportional to the square of the number of particles, making the complexity of the algorithm  $\mathcal{O}(N^2)$ \cite{luijten1995, fukui2009}.

Another challenge involves generating uncorrelated (independent) samples near the critical point, which are necessary for calculating the desired quantities. Close to the critical point, the system dynamics significantly decelerate due to the phenomenon of critical slowing down. This results in a gradual evolution over time and a high correlation between samples in a sequence. As such, the efficiency of an algorithm is determined by its ability to produce a larger number of uncorrelated samples in a shorter computational time \cite{newman1999, janke2008, flores-sola2017}. 

A variety of methods are available to solve these issues for particular systems, with the most celebrated ones being cluster-update methods for the short-range and long-range Ising model. For the short-range Ising model, the Swendsen-Wang \cite{swendsen1987} and Wolff \cite{wolff1989} methods activate bonds between spins to form large clusters and then flip the spins inside the clusters \cite{newman1999}. These methods effectively eliminate critical slowing down in the short-range Ising spin model \cite{janke2008, newman1999}. However, for the long-range counterpart, they result in $\mathcal{O}({N^2})$ complexity since each bond must be checked for activation \cite{luijten1995}. 

To address this issue in cluster-update methods for long-range interactions, Luijten and Bl\"{o}te took a different approach. They identified the activated bonds using the cumulative distribution of the bond activation probability. This approach effectively reduces the complexity to $\mathcal{O}(N \ln N)$  \cite{luijten1995}. Later, Fukui and Todo improved this cluster-update method by treating bond activations as distributing Poisson events on bonds, which can be done with $\mathcal{O}(N)$ complexity \cite{fukui2009}. Cluster-update methods can also be employed to simulate the isotropic $n$-vector spin models (\textit{i.e.}, Heisenberg and XY models) by projecting the spins onto random directions and treating the projected spins as Ising variables \cite{newman1999, luijten1995}. Unfortunately, cluster-update methods cannot be used to simulate magnetic dipolar systems because, unlike $n$-vector spin models, the dipolar interaction is anisotropic \cite{baek2011}.

Over the past two decades, irreversible MC methods have garnered attention due to their superior efficiency in specific scenarios when compared to the Metropolis algorithm \cite{diaconis2000, faizi2020, turitsyn2011, ottobre2016, suwa2010, ichiki2013, ren2006}. The event-chain MC method is a powerful technique for reducing critical slowing down in systems with continuous degrees of freedom and isotropic interactions. This method is effective for both short-range and long-range interactions, especially for central force systems and $n$-vector spin models \cite{bernard2009,michel2014,michel2015,nishikawa2015}. For long-range interactions, the event-chain method can reduce the complexity to $\mathcal{O}(N)$ by employing thinning methods \cite{kapfer2016}

In an event chain simulation of an $n$-vector spin model, a spin undergoes a continuous rotation until a lifting event occurs, transferring the rotation to another spin \cite{michel2015}. However, in spin systems with anisotropic interactions such as dipolar interactions, a lifting event can result in the rotation of multiple spins, which complicates the computation \cite{nishikawa2016}. Although the issue of multiple spin rotation has been addressed for many-body interactions \cite{harland2017}, it remains a problem for systems with dipolar interactions \cite{hollmer2022}.

The clock \cite{michel2019}, stochastic cutoff (SCO) \cite{sasaki2008}, and Tomita \cite{tomita2009} Monte Carlo methods reduce the computational complexity of the simulation of long-range interacting systems to $\mathcal{O}(N)$. These methods are particularly effective for simulating two-dimensional dipolar systems. Both the SCO and Tomita methods have been specifically tested and proven to work well for these types of systems \cite{tomita2009, sasaki2008, komatsu2018}.

While these three methods are based on different ideas, they all share a common approach to reducing computational complexity.
This approach is to evaluate pairwise interactions for only a subset of spins that are randomly selected, primarily from those with stronger interactions. Techniques like the dynamic thinning or Fukui-Todo methods can be used for this random selection process \cite{michel2019, sasaki2008, tomita2009}. Due to this similarity, one might ask which of these methods is most efficient for simulating 2D dipolar systems.

In their simplest form, clock, SCO, and Tomita methods have limitations, including lower acceptance rates compared to the Metropolis method \cite{michel2019, muller2023}, and increased computational time due to the extensive generation of random numbers \cite{tomita2016, komatsu2018}. These issues can be addressed in the clock method by grouping interactions into boxes of tunable sizes \cite{michel2019}. The suggested boxing strategy is to group interactions of similar strength into boxes of constant size, where the size of the boxes is determined by the type of interactions \cite{michel2019}. The concept of tunable boxes is theoretically available in the SCO and Tomita methods, but it is not explicitly stated in the literature. In these methods the grouping of only near neighbors is used, which is reported to be effective for 2D dipolar systems \cite{tomita2016, sasaki2008, komatsu2018}.  Nevertheless, to the best of the authors' knowledge, the impact of the near-neighbors box size on the acceptance rate and computational time of the simulation has not been extensively studied.

In this work, we compared the clock, SCO, and Tomita methods for simulating a triangular lattice of pure dipoles near the critical point. We investigated the effect of different near-neighbors box sizes on the acceptance rate and computational time of these methods. Additionally, we studied the addition of overrelaxation moves to these methods,  which improved the performance of the simulations. We demonstrate that the efficiency of these methods significantly improves when using an appropriate box size and incorporating overrelaxation moves, making them much more suitable for simulating dipolar systems near the critical point. The efficiency of these methods becomes similar to each other when employing the near-neighbors box and overrelaxation moves. Therefore, the choice of the best algorithm boils down to its simplicity in the theory and implementation.

\section{\label{sec:discussion}Hamiltonian} 
The Hamiltonian for a two-dimensional lattice of spins with both the exchange and dipolar interactions in free boundaries is given by
\begin{equation} 
\mathcal{H} = -J\sum_{\langle i,j \rangle}\vec{S}_i\cdot\vec{S}_j - \frac{1}{2} \frac{\mu_0}{4\pi} \sum_{\substack{i, j \\ i \ne j}}\frac{ 3(\vec{S}_i\cdot\vechat{r}_{ij})(\vec{S}_j\cdot\vechat{r}_{ij}) - \vec{S}_i\cdot\vec{S}_j}{|\vec{r}_{ij}|^3}. \label{eq:hamiltonian}
\end{equation}
Here, $\vec{S}_i$ represents the spin vector at site $i$. The first term denotes the exchange interaction between neighboring spins, with $J$ being the exchange coupling constant. The notation $\langle i,j \rangle$ indicates that the sum is taken over all pairs of nearest neighbors. The second term corresponds to the dipolar interaction between all pairs of spins, where $\mu_0$ is the vacuum permeability. The vector that represents the relative position of spin $j$ from spin $i$ is denoted by  $\vec{r}_{ij} = |\vec{r}_{ij}| \, \vechat{r}_{ij}$. The factor $1/2$ is due to double counting.

To express the Hamiltonian in a more concise form, we define the dipole interaction tensor $\mathbf{D}_{ij}$ using
\begin{equation} \label{eq:dyadic}
\vec{D}_{ij} = \frac{\mu_0}{4\pi|\vec{r}_{ij}|^3}\left(3\vechat{r}_{ij}\vechat{r}_{ij} - \mathbf{I}\right),
\end{equation}
in the case of $i \ne j$. Here $\vechat{r}_{ij}\vechat{r}_{ij}$ is a dyadic tensor, and $\mathbf{I}$ is the identity tensor. Since a spin does not interact with itself, here we have $\vec{D}_{ii} = \mathbf{0}$. Note that the dipolar tensor is both symmetric and anisotropic. This is because the property $\vec{S}_i \cdot \vec{D}_{ij} \cdot \vec{S}_j = \vec{S}_j \cdot \vec{D}_{ij} \cdot \vec{S}_i$ ensures symmetry, while it is anisotropic since the dipolar interaction depends on the orientation of $\vec{r}_{ij}$. The Hamiltonian of a magnetic system with the exchange and dipolar parts can be reformulated in terms of dipole interaction tensors $\vec{D}_{ij}$ as 
\begin{equation}\label{eq:hamiltonianSimple}
\mathcal{H}(\{\vec{S}_i\}) = -J\sum_{\langle i,j \rangle}\vec{S}_i\cdot\vec{S}_j - \frac{1}{2} \sum_{i, j} \vec{S}_i \cdot \vec{D}_{ij} \cdot \vec{S}_j.
\end{equation}
Besides, the exchange interaction can be absorbed in $\vec{D}_{ij}$ by addition of $J \vec{I}$ to $\vec{D}_{ij}$ for $j \in \mathrm{n.n.}(i)$ . Since we are interested in purely dipolar systems, we set $J=0$ throughout this article.
The dyadic tensors contain only the information regarding the location of spins. Therefore, with this notation, the effects of locations in the dipole-dipole interaction are separated from the orientations of spins. As the particles in a magnetic material have fixed positions, these tensors remain constant throughout the simulation. Consequently, they only need to be computed once at the start of the simulation, which help save a lot of computing power \cite{girotto2018}.

In systems with periodic boundary conditions, a spin interacts not only with other spins but also with their periodic images. To simplify the description, such a system can be imagined as being placed in a supercell, which repeats periodically in space. For systems with only short-range interactions, it is sufficient to consider interactions with periodic images in adjacent supercells (\textit{i.e.}, replicas of the system). However, in systems with long-range interactions, it is necessary to account for interactions with periodic images in all the supercells. 

We denote the basis vectors of the supercell as $\vec{a}$ and $\vec{b}$. For instance, in a square supercell with a side length $L$, the basis vectors would be $\vec{a} = L \vechat{x}$ and $\vec{b} = L \vechat{y}$. We also denote translation vectors to other supercells using $\vec{R}_{\mu\nu} = \mu \vec{a} + \nu \vec{b}$. 

To account for the interaction with all the periodic images, we redefine the dipole interaction tensor in this case as follows:
\begin{equation} \label{eq:dyadicPeriodic}
\vec{D}_{ij} = \sideset{}{'}\sum_{\mu, \nu = -\infty}^{+\infty} \frac{\mu_0}{4\pi|\vec{r}_{\mu\nu}|^3}\left(3\vechat{r}_{\mu\nu}\vechat{r}_{\mu\nu} - \mathbf{I}\right),
\end{equation}
where we defined the relative position vector from spin $i$ to the periodic image of spin $j$, located in a supercell at $\vec{R}_{\mu\nu}$ as
\begin{equation} \label{eq:rmunu}
\vec{r}_{\mu\nu} = \vec{r}_{ij} + \vec{R}_{\mu\nu}.
\end{equation}
To write this simplified notation for relative position vectors, we used the fact that the spins $i$ and $j$ are fixed when calculating the tensor $\vec{D}_{ij}$. The prime in the summation indicates that when $i=j$, the case of $\mu=\nu=0$ is excluded. Therefore, in $\vec{D}_{ii}$, we only take into account the interaction of the spin with its periodic images. As the summation in Eq. \eqref{eq:dyadicPeriodic} converges slowly, it is usually computed through the Ewald summation technique \cite{weis2003, debell2000}.

The Hamiltonian in the periodic boundary condition can be written as
\begin{align}% \label{eq:hamiltonianPeridic}
\mathcal{H}(\{ \vec{S}_i \}) &= - \frac{1}{2} \sum_{i,j} \vec{S}_i \cdot \vec{D}_{ij} \cdot \vec{S}_j\nonumber\\ \label{eq:hamiltonianPeridicFields}
&= - \frac{1}{2} \sum_i  \vec{S}_i \cdot \vec{H}_{i}  - \sum_i \vec{S}_i \cdot \vec{D}_{ii} \cdot \vec{S}_i. 
\end{align}
Here, $\vec{H}_{i}$ is the effective field on spin $\vec{S}_{i}$ and is defined as 
\begin{equation}\label{eq:effectivefield}
\vec{H}_{i} = \sum_{j(\ne i)} \vec{D}_{ij} \cdot \vec{S}_j.
\end{equation}
In the particular case of the square and triangular lattices, $\vec{D}_{ii} $ is proportional to the identity tensor. Therefore, the self-interactions are constants and can be omitted from Eq. \eqref{eq:hamiltonianPeridicFields}. 
\section{Monte Carlo Methods and the Metropolis algorithm \label{sec:Metropolis}}
The Metropolis algorithm is a powerful tool for simulating the behavior of complex systems. In the context of a continuous spin system, it generates a sequence of samples that follow the Boltzmann distribution. This is achieved through a process of proposing and accepting new spin configurations based on their energy difference from the current configuration. Algorithm~\ref{alg:metropolis} presents the Metropolis algorithm with sequential updating applied to a continuous spin system \cite{ren2006,metropolis2004,newman1999}.
\begin{algorithm}[H]
\caption{Traditional Metropolis with sequential updating}\label{alg:metropolis}
\begin{algorithmic}[1]
\For{spins $1$ to $N$}
	\State Candidate $\vec{S}_{i}' \gets$  a random direction
	\State Accept $\vec{S}_{i} \gets \vec{S}_{i}'$ with probability $P = e^{-\beta [\Delta E]^{+}}$
\EndFor
\end{algorithmic} 
\end{algorithm}
In the 3\textsuperscript{rd} line of the algorithm, $[\Delta E]^{+}$ is defined as 
\begin{equation}
[\Delta E]^{+} = \mathrm{max}(\Delta E, 0),
\end{equation}
and $\Delta E$ represents the energy difference between the new and the current spin configuration. The operator $[\phantom{\Delta}]^+$ makes the transition probability equal to one when $\Delta E < 0$. The coefficient $\beta$ represents the inverse of temperature. The probability
\begin{equation}\label{eq:MetropolisFilter}
P = e^{-\beta [\Delta E]^{+}}
\end{equation}
is sometimes referred to as the Metropolis filter \cite{michel2015,michel2019}.

Two essential problems arise when using the Metropolis algorithm to simulate a system of dipoles near the critical point. The first is that the calculation of $\Delta E$ to change $\vec{S}_i$ to $\vec{S}'_i$ in a long-range interaction system,
\begin{equation}
\Delta E = - (\vec{S}'_i - \vec{S}_i) \cdot \sum_{j(\ne i)} \vec{D}_{ij}  \cdot \vec{S}_j
\end{equation}
is computationally expensive. Because every spin interacts with all $N$ other spins, the calculation is an $\mathcal{O}(N)$ operation. Therefore, the complexity of an MC step (sweep) is $\mathcal{O}(N^2)$ or $\mathcal{O}(L^4)$ in a two-dimensional $L \times L$ lattice. Another way to calculate the energy difference is to use the effective field on spin $i$ as
\begin{equation}
\Delta E = - (\vec{S}'_i - \vec{S}_i) \cdot \vec{H}_i. 
\end{equation}    
This operation is $\mathcal{O}(1)$ if we had precalculated $\vec{H}_i$ for all spins. So, if the MC move is rejected all we have done is an $\mathcal{O}(1)$ operation. But if it succeeds, we need to perform an $\mathcal{O}(N)$ operation to update the effective field on all other spins  as
\begin{equation}
\vec{H}_j \gets \vec{H}_j + \vec{D}_{ij} \cdot (\vec{S}'_i - \vec{S}_i). \label{eq:updateeffectivefield}
\end{equation} 
So, although the complexity is still $\mathcal{O}(N^2)$ using the effective fields method, it is reduced by the constant factor of the acceptance ratio $P_\mathrm{acc}$ \cite{hucht1995}.

The other challenge with the Metropolis algorithm for simulating a system near the critical point is the increased statistical error. In an MC simulation, a new sample is generated from the previous sample. This leads to a correlation between sequential samples, which increases the error in statistical averaging results. The degree of correlation depends on the algorithm used. In the case of the Metropolis algorithm, this correlation is significant near the critical point \cite{janke2008}.

In an experiment where a quantity $Q$ is measured $n$ times, the best estimate of the quantity is the average of those $n$ measurements,
\begin{equation}\label{eq:Qbar}
\overline{Q} = \frac{1}{n} \sum_{i=1}^{n} Q_i,
\end{equation}
where  $\overline{Q}$  is the average of measurements and $Q_i$ represents the $i^{\text{th}}$ measurement of the quantity $Q$. The best estimate of the error is given by
\begin{equation}\label{eq:error}
\mathrm{err} \simeq \frac{\sigma}{\sqrt{n}} \simeq \sqrt{\frac{\overline{Q^2} - \overline{Q}^2}{n}},
\end{equation}
where $\sigma$ is the standard deviation of the measurements. 

In Eq. \eqref{eq:error}, it is assumed that the measurements $Q_i$ are statistically independent. However, as we noted before, the generated samples are correlated in an MC simulation. The correlation in quantity $Q$ can be computed using the autocorrelation function of $Q$, which is defined as
\begin{equation}\label{eq:phit}
\phi_Q(t) = \frac{\langle Q(0) Q(t) \rangle - \langle Q(0) \rangle \langle Q(t) \rangle}{\langle Q^2(0) \rangle - \langle Q(0) \rangle^2},
\end{equation}
where the time $t$ is measured in MC steps and $ \langle \, \cdots \, \rangle$ denotes the ensemble average. As the MC time $t$ increases from a given initial sample, new samples become less correlated with the first sample. To find the time interval where two measurements of $Q$  are statistically uncorrelated, one can use the autocorrelation time of the quantity $Q$ which is given by 
\begin{equation}\label{eq:tau}
\tau_{Q} = \frac{1}{2} + \sum_{t=1}^{\infty} \phi(t).
\end{equation}
The autocorrelation time of the algorithm $\tau$ is defined as $\tau=\max(\{\tau_{Q}\})$, \textit{i.e.} autocorrelation time of the slowest variable in the algorithm \cite{newman1999}.

It can be shown that the samples that are separated by intervals of $2\tau$ are approximately statistically independent \cite{newman1999, janke2008}. This implies that, in an MC simulation, the effective number of measurements in Eq. \eqref{eq:error} can be estimated by 
\begin{equation}\label{eq:neff}
n_\mathrm{eff} \simeq \frac{t_\mathrm{max}}{2\tau},
\end{equation}
where $t_\mathrm{max}$ is the length of the MC simulation. Therefore, a shorter autocorrelation time leads to a greater number of independent samples, which in turn improves the accuracy of statistical averaging.

As a finite-sized system gets closer to the critical point, the autocorrelation time increases significantly, following the equation
\begin{equation}\label{eq:dynamicalExp}
\tau \propto L^{z}.
\end{equation}
Here, $z$ is the dynamical exponent and depends on the algorithm being used \cite{janke2008, newman1999}. 
For the Metropolis algorithm, $z \simeq 2$ \cite{newman1999}, resulting in a rapid increase of the autocorrelation time with the lattice size. Hence, a considerable amount of time must be spent in terms of the MC step to reach an independent sample, a phenomenon known as critical slowing down.
For some systems, efficient algorithms can drastically reduce the dynamical exponent.  For example in 1D long-range Ising model the cluster-update method using the Fukui-Todo approach reduces the dynamical exponent to  $z \simeq 0.5$ \cite{fukui2009, flores-sola2017} and the event-chain method shows $z \simeq 1$ for the 3D long-range Heisenberg model \cite{nishikawa2015}. Unfortunately, both of these algorithms are not applicable for dipolar systems due to the anisotropy of dipole-dipole interaction \cite{baek2011a, hollmer2022}.

\section{Efficiency of Algorithms}
In the previous section, we discussed the Metropolis algorithm and pointed out its shortcomings in simulating dipolar systems. To quantitatively assess the performance of an MC algorithm, we define the effective computational time. This parameter is the computational time required to generate an uncorrelated sample and serves as the measure of algorithm efficiency  \cite{flores-sola2017, elci2013}.  Because after generating approximately $2\tau$ samples, the last sample is uncorrelated with the initial sample, we define the effective computational time as
\begin{align}
t_{\mathrm{eff}} &= 2\tau \times t_{\mathrm{MCS}} \label{eq:teff} \\
 &\sim L^z \times t_{\mathrm{MCS}}. \notag
\end{align}
Here, $t_{\mathrm{MCS}}$ is the single ``MC Step" time (\textit{i.e.}, the computational time of generating a sample).
When considering the Metropolis algorithm for simulating a dipolar system, the computational time $t_{\mathrm{MCS}}$ increases with the system length by scale of $L^4$, and the dynamical exponent takes the value of $z \simeq 2$. Consequently, the effective computational time scales as $L^2 \times L^4 \sim L^6$.

When examining a dipolar system near its critical point, the inefficiency of the Metropolis algorithm becomes particularly noticeable. This is attributed to the fact that the investigation of critical behavior often necessitates the simulation of large-scale systems. Consequently, alternative algorithms are required that can perform these tasks in a shorter amount of time.

\section{Overrelaxation move}
\begin{figure}[tbp] % fig 1
	\centering
	\includegraphics[width=0.5\columnwidth]{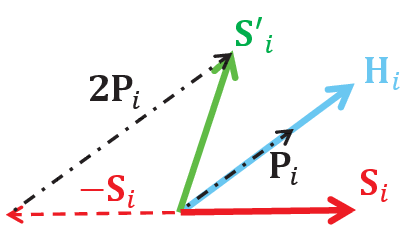}
	\caption{\label{fig:overrelax} The vector $\vec{P}_i$ represents the projection of $\vec{S}_i$ onto $\vec{H}_i$. The vector $\vec{S}'_i$ is the reflection of $\vec{S}_i$ about $\vec{H}_i$, and it can be calculated using $\vec{S}'_i = -\vec{S}_i + 2\vec{P}_i$.} 
\end{figure}

When proposing to change the spin $\vec{S}_i$ to $\vec{S}'_i$ in the Metropolis method, it is common to choose $\vec{S}'_i$ to point in a random direction. We call this a \emph{random move} which is a small random displacement of the system in phase space. However, in order to reduce the autocorrelation time, it is more efficient to make this displacement as large as possible. A straightforward yet effective method for accomplishing this task is to use an \emph{overrelaxation move} \cite{creutz1987}.
In the Metropolis Method for spin systems, the overrelaxation move involves choosing the new trial spin by reflecting the old spin about the effective field vector (see Fig.~\ref{fig:overrelax}). This can be represented mathematically as
\begin{equation} \label{eq:Sioverrelax}
\vec{S}'_i = -\vec{S}_i + 2 \left( \frac{\vec{S}_i \cdot \vec{H}_i}{H_i^2} \right)  \vec{H}_i.
\end{equation}

In the absence of dipolar interaction, an overrelaxation move in the Metropolis algorithm is rejection-free and keeps the energy constant. So Metropolis algorithm is done with a combination of random and overrelaxation moves to ensure ergodicity.

When the energy is kept constant through an overrelaxation move, it often has the highest effect in reducing the autocorrelation time, which can even decrease the dynamical exponent to $z \simeq 1$. It is common practice to choose an overrelaxation to random move ratio of approximately 10:1, which results in a dynamical exponent of $z \simeq 1.5$ \cite{peczak1993, pixley2008, gvozdikova2005}. In this work, we chose a 10 to 1 overrelaxation to random move ratio, unless otherwise strictly specified.

Therefore, the addition of the overrelaxation moves to the Metropolis algorithm, typically, leads to an effective time of approximately  $t_{\mathrm{eff}} \sim L^{5.5}$, which is slightly smaller than $t_{\mathrm{eff}} \sim L^{6}$ for the Metropolis algorithm with only random moves.

When the Hamiltonian is such that the energy can not be kept constant in an overrelaxation move, the autocorrelation time will not be greatly reduced and the dynamical exponent remains $z \simeq 2$. This case happens in a dipolar system with the periodic boundary condition because of the self-energy of a spin with its periodic images. Only in some symmetries such as the triangular and square lattices the energy can be kept constant since the self-energy contributes to a constant in the Hamiltonian \cite{stasiak2009}.

\section{Monte Carlo methods with $\mathcal{O}(N)$ complexity}
This study aimed to identify the most suitable algorithm for simulating a 2D lattice of pure dipoles near the critical point using the MC method. Various MC algorithms with $\mathcal{O}(N)$ complexity have been proposed in the scientific literature for long-range interactions, which are capable of simulating dipolar interacting systems on lattices \cite{michel2019, sasaki2008, tomita2009, muller2023}. We have reviewed three methods: the clock, stochastic cutoff (SCO), and Tomita. These methods rely on a common approach to reducing complexity, and we tried to identify the best of them for simulating the mentioned system. Additionally, we have developed a code that can utilize any one of these three methods  \cite{onmc}. Since the clock method is the simplest to explain, we will provide a detailed explanation of it here, while the SCO and Tomita methods will be covered briefly.

\subsection{Clock method \label{sec:clock}}
In the Metropolis algorithm, a modification is proposed to the spin at a certain position $i$. This suggested change is then either accepted or rejected based on the Metropolis filter, $P = e^{-\beta[\Delta E]^{+}}$.
Calculating the energy change $\Delta E$ is computationally expensive. This is because the dipolar interaction is long-range, so the energy change of the $i^\text{th}$ spin with all of its $N-1$ neighbors needs to be computed.

Alternatively, in the clock method, this change in the spin $i$ is accepted by the factorized Metropolis filter as \cite{michel2014}
\begin{equation}\label{eq:fac}
P = \prod_{j} e^{-\beta[\Delta E_j]^{+}} = \prod_{j} P_j.
\end{equation}
Here, $j$ represents the indices for the neighbors of $i^\text{th}$ spin. $\Delta E_j$ shows the difference in the interaction energy between spins $i$ and $j$, before and after the change. $P_j$ denotes the probability that the neighbor $j$ accepts the change. In other words, the total probability is decomposed into a product sequence, where each component depends only on one of the neighbors. We adopt a simplified notation where $\Delta E_j$ and $P_j$ appear independent of the index $i$. However, it is important to note that both quantities depend on $i$ in the full expression. This simplification is feasible because $i$ denotes the changing spin, which remains constant during the sampling process of the factorized Metropolis filter.

Accepting or rejecting the change by directly computing the factorized metropolis filter is an operation of order $N$. Because one needs to calculate $P_j$ for all the neighbors.

Another approach is to break the factorized Metropolis filter into $N - 1$ consecutive trials, each occurring with probability $P_j$. In order to accept the change, all the trials must be successful. If any of the trials fail, it means that the change is completely rejected. To minimize the number of trials evaluated, we can begin with the nearest neighbors since they have a higher chance of failure. By starting with these neighbors, we can identify rejections early in the process, significantly reducing the computational workload compared to evaluating all $N - 1$ trials.

Since these trials are performed sequentially,
The first trial in the factorized Metropolis filter that does not succeed, is defined as  ``the first failure event" \cite{michel2019}. When this event occurs, regardless of the results of the next trials, the change is rejected. The probability of the first failure event to happen in the $j^\mathrm{th}$ factor is given by
\begin{equation}
P_{\text{rej}}(j) = h_j \prod_{k=1}^{j-1}(1-h_k) \label{eq:rej}.
\end{equation}
Here, $h_j = 1-P_j$ represents the probability of failure in the $j^\mathrm{th}$ trial \cite{fan2023}. Acceptance occurs when all trials are successful. In this case, the probability of acceptance is
\begin{equation}
P_{\text{acc}} = P = \prod_{k=1}^{N-1}(1-h_k) \label{eq:acc}.
\end{equation}

These events naturally form a probability distribution function (PDF) of size $N$ \cite{michel2019}, denoted as $\mathcal{P}(j)$:
\begin{equation}\label{eq:distribution}
\mathcal{P}(j) = \begin{cases} P_{\text{rej}}(j) & \text{if} \ 1 \leq j \leq N-1, \\ P_{\text{acc}} & \text{if} \ j = N. \end{cases}
\end{equation}
To determine the fate of the change, it is sufficient to sample from $\mathcal{P}$. If $j = N$, the change is accepted. Otherwise, the change is rejected by the $j^\mathrm{th}$ factor. The challenge lies in efficient sampling from $\mathcal{P}$, since efficient methods such as the ``Walker alias" or ``inversion sampling" require the PDF to be fixed during the simulation \cite{walker1977, newman1999}. However, $\{h_j\}$ depend on the arrangement of spins and vary throughout the simulation, making the PDF configuration-dependent. 

\begin{figure}[tbp] % fig 2
    \centering
	\includegraphics[width=\linewidth]{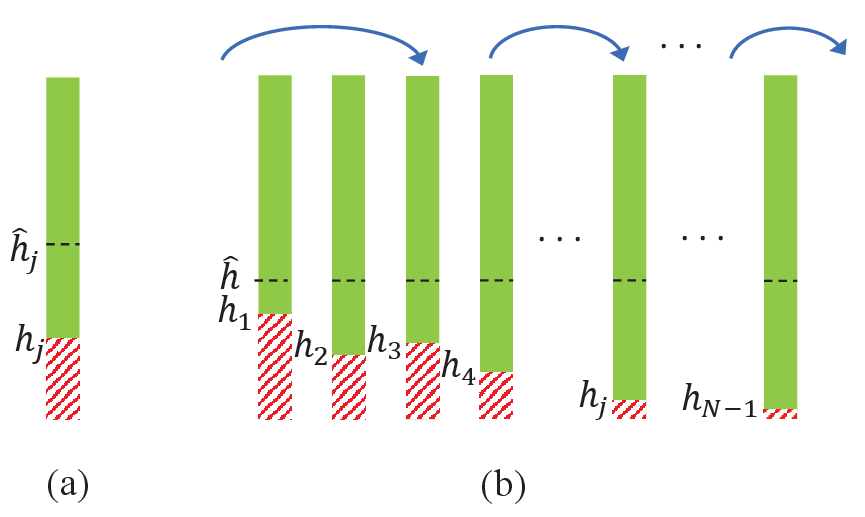}
    \caption{%
	(a) Probability diagram illustrating success and failure in the $j^\text{th}$ trial. In this diagram, the shaded region (\shaded) represents the probability of success, while the hatched region ({\protect\hatched}) indicates the probability of failure.  The portion of the column below the horizontal dashed line signifies the bound failure probability. (b) Sequential execution of $N$ trials, where the bound failure probability for all factors is set to the maximum failure probability of the nearest neighbor, \textit{i.e.}, $\hat{h} = \max({h}_1)$. A typical jumping through bound failure events is shown using arrows.}\label{fig:clock} 
\end{figure}

To address this issue, we focus on the observation that the interaction of spin $i$ with its distant neighbors is very weak and the maximum failure probability of the trial in a distant neighbor is extremely low. Therefore, we can define ``the bound failure probability" as $\hat{h}_j \geq \max(h_j)$ such that $\hat{h}_j$ is not dependent on the configuration of spins. The maximum failure probability $\max(h_j)$  occurs when the difference in the interaction energy $\Delta E_j$ between spins $i$ and $j$  is the highest among all possible configurations \cite{michel2019}. For instance, consider a pair of spins that interact only through exchange interaction. The maximum energy difference arises when the spins transition from a configuration in which they are aligned to one where they are oppositely aligned. 

To reduce the complexity of sampling from $\mathcal{P}$, we define the bound trial which fails with the bound failure probability $\hat{h}_j$. Fig.~\ref{fig:clock}(a), schematically shows the failure probability ($h_j$), success probability ($P_j$) and bound failure probability ($\hat{h}_j$) for a typical case.

If the result of sampling from the bound trial is successful, it means that the actual trial has definitely succeeded. However, if the bound trial fails, the result of the actual trial is inconclusive. In this case, we need to sample a verification trial to determine the fate of the actual trial \cite{michel2019}. This verification trial fails with the conditional probability of
\begin{equation}\label{eq:hverification}
h_{j,\text{ver}} = \frac{h_j}{\hat{h}_j}.
\end{equation}

If the sampling from the verification trial results in a failure, then the actual trial fails; otherwise, it is a success. In Fig.~\ref{fig:clock}(a) the conditional probability can be interpreted as the likelihood of being situated within the hatched area while already located below the dashed line.

A straightforward method for sampling $\mathcal{P}$ is to assign the maximum failure probability of the nearest neighbor, which possesses the strongest interaction, as the bound failure probability for all neighbors. In other words, we set $\hat{h}_k = \hat{h} = \max(h_1)$ for all $k$; see Fig.~\subfigref{fig:clock}{b}. Consequently, the PDF of the first failure event in the bound trials follows a geometric distribution,
\begin{equation}\label{eq:geometric_distribution}
\mathcal{P}(j) = \begin{cases} \hat{h}(1-\hat{h})^{j-1} & \text{if} \ 1 \leq j \leq N-1, \\ (1-\hat{h})^{N} & \text{if} \ j = N. \end{cases}
\end{equation}
Using the inverse transform sampling, the first failure event of bound trials can be found by inverting the PDF as
\begin{equation}\label{eq:geometric_rejection}
j = \left\lceil \frac{\log(1-u)}{\log \big(1-\hat{h} \big)} \right\rceil.
\end{equation}
In this equation, $u$ is a uniform random number in the interval $[0,1)$, and the notation $\lceil \phantom{i} \rceil$ represents the ceiling function. Since $j$ shows the index of the first bound failure event, we subsequently sample the verification trial using Eq. \eqref{eq:hverification} to confirm whether it results in a failure in the actual trial or not. If it is an actual failure, then the change is rejected. Otherwise, it is a false failure event and we jump to the next failure event of bound trials by \cite{michel2019}
\begin{equation}\label{eq:next_rejection}
j = j' + \left\lceil \frac{\log(1-u)}{\log \big( 1-\hat{h} \big)} \right\rceil,
\end{equation}
where $j'$ indicates the index of the last failure event. Then, we verify if it is an actual failure. We repeat this process until $j \geq N$. If the change is not rejected by this point, it is accepted. Fig \ref{fig:clock}(b) shows a typical jumping through bounds failure events. 	

This process could be optimized using the dynamic thinning method by lowering the bound failure probabilities following a false failure event \cite{shanthikumar1985, michel2019}. Specifically, we set $\hat{h} = \max(h_{j'+1})$, where $j'$ is the index of the false failure event. The other way is to use the Fukui-Todo method, which utilizes Poisson processes to sample the probability distribution \cite{michel2019, fukui2009}. Detailed explanations of these methods can be found in Appendices~\ref{app:dythin} and~\ref{app:fukui}.

\begin{figure}[tb] % fig 3
	\centering
	\includegraphics[width=\linewidth]{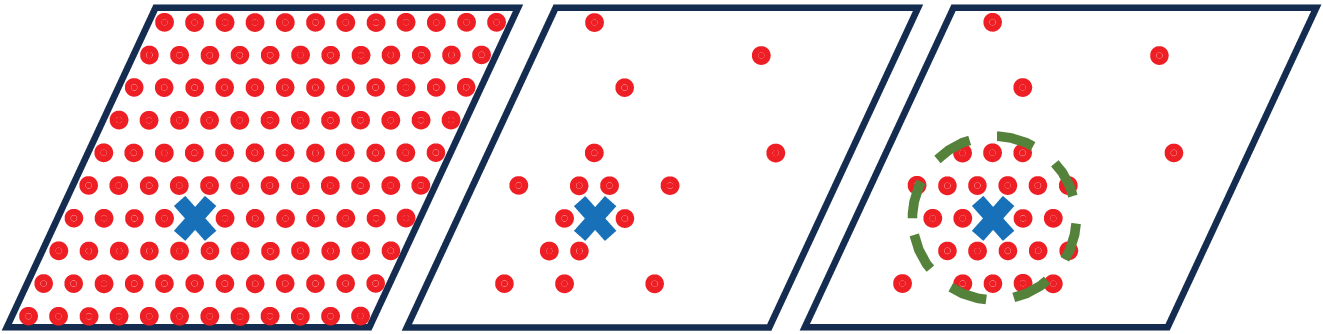}
	\caption{\label{fig:updaterulebox} (left) The Metropolis method computes the total energy difference of the changing spins with all of its neighbors. (middle): The clock, SCO, and Tomita methods compute the energy difference with only some of the neighbors, chosen randomly based on the interaction strength with the changing spin. (right): Close neighbors are grouped into a box (dashed circle) for the clock, SCO, and Tomita methods. This approach reduces computational time by limiting the generation of random numbers, required for stochastic selection of neighbors. It also increases the acceptance rate by compensating for the energy difference [see Eq. \eqref{eq:MetropolisFacinequality}]. The changing spin is indicated by the symbol (\textcolor[rgb]{0.211, 0.423, 0.745}{$\bm\times$}), and filled circles (\textcolor[rgb]{0.929, 0.113, 0.141}{$\bullet$}) are used to denote neighbors that require pairwise interaction evaluation.}
\end{figure}

The biggest problem with the clock algorithm is the lower acceptance rate compared to the Metropolis algorithm. This is due to the compensation of the positive and negative $\Delta E_{j}$ terms in the Metropolis algorithm, as described by the following inequality,
\begin{equation}\label{eq:MetropolisFacinequality}
\left[\sum_{j} \Delta E_{j}\right]^{+} \leq \sum_{j}\left[\Delta E_{j}\right]^{+}.
\end{equation}

This lower acceptance rate causes the state of the system to evolve more slowly, resulting in a longer autocorrelation time \cite{michel2019}. The acceptance rate can be increased by utilizing the factorized Metropolis filter with the boxes technique \cite{michel2019}. In this approach, instead of factoring each individual neighbor, they are grouped into sets of tunable size as
\begin{equation}\label{eq:pfacbox}
P =\prod_{k=1}^{N_{\mathrm{B}}} \exp \left(-\beta \left[\sum_{j \in B_k}  \Delta E_{j}\right]^{+}\right).
\end{equation}
Here, $N_\mathrm{B}$ represents the number of boxes and $B_k$ represents the $k^\text{th}$ box. According to Eq. \eqref{eq:MetropolisFacinequality}, by grouping the interactions, the failure probability of a box lowers compared to considering the neighbors as individuals.

In strictly extensive systems, such as 2D dipolar lattices, the interactions are of the order \( r^{-\sigma} \), with \( \sigma > d \) where \( d \) is the dimension of the system. For these systems, the state of a single spin can be updated with \( \mathcal{O}(1) \) complexity using the dynamic thinning or Fukui-Todo methods \cite{michel2019}. Since all $N$ spins of the system must be updated in an MC step, the computational time of the clock method for an MC step is $\mathcal{O}(N)$, or $\mathcal{O}(L^2)$ in an $L\times L$ lattice. However, the clock method's complexity is not as straightforward. It depends on the sharpness of the bound and the used boxing. For details refer to Ref. \cite{michel2019}.

\subsection{Stochastic cutoff and Tomita methods}
In the Metropolis algorithm, the acceptance of a move depends on the total energy difference between the changing spin and all of its neighbors [see Fig.~\spacedsubfigref{fig:updaterulebox}{left}].
Because the interaction with close neighbors is stronger, they have a greater influence on the decision to reject or accept the move, while distant neighbors exert minimal influence on this decision.
This fact is utilized in the clock method, where a change is accepted only if all neighbors individually accept it. Since interactions with distant neighbors are weak, they almost always accept the change \cite{michel2019}. Therefore, the interaction energy is evaluated with only a small number of randomly selected spins, which are mostly from the closer neighbors, as shown in Fig.~\spacedsubfigref{fig:updaterulebox}{middle}.

The property of minimal influence from distant neighbors is also employed in the stochastic cutoff (SCO) and Tomita methods, each with a distinct conceptual approach \cite{sasaki2008, tomita2009}. In the SCO method, interactions are stochastically switched to 0 with a probability based on their interaction energy with the changing spin. Since the interactions with distant spins are weak, most of them will switch to 0. The switching procedure is similar to that in the clock method, where we jump through bound rejections according to Eq. \eqref{eq:next_rejection}, or more efficiently, using the dynamic thinning or Fukui-Todo methods. Switching interactions comes with the cost that, when an interaction fails to switch, it will be changed to a pseudopotential; for details, see Appendix~\ref{app:SCO}. Then, based on the total energy difference $\Delta E$ between that changing spin and the failed-to-switch interactions (pseudopotentials), the move is decided to be either accepted or rejected \cite{sasaki2008}.

The Tomita method was originally developed for the long-range Ising spin model based on the cluster-update method and the Fukui-Todo complexity reduction approach. In this method, the Boltzmann weight of a configuration is stochastically adjusted so that interactions with most of the distant neighbors are removed. As this method is more sophisticated than the clock and SCO methods, it will not be explained in this article. Interested readers could refer to its original articles for the implementation details \cite{tomita2009a, tomita2009, tomita2016}.

The concept of tunable boxes in the clock method can be extended to the SCO and Tomita methods as well. In the SCO method, the simplest form of ``box'' is to exclude some of the spins from the switching process. The interaction potentials for these excluded spins do not change to pseudopotentials. Despite that it is not essentially a box, we refer to these excluded spins as being put into a \emph{box}.  For the real tunable boxes in the SCO method, similar to the one described in Eq. \eqref{eq:pfacbox} for the clock method \cite{michel2019}, refer to Appendix~\ref{app:SCO}.

\subsection{Boxing near neighbors}
Although the clock, SCO, and Tomita methods reduce the complexity of simulating a triangular lattice of dipoles to $\mathcal{O}(N)$, if they are used without the boxing technique they can have lower acceptance rates compared to the Metropolis method \cite{michel2019, komatsu2018} and long computational times \cite{tomita2016, sasaki2008}. Our results demonstrate that these methods only outperform the Metropolis algorithm (with overrelaxation moves) for systems with lengths exceeding  $L \approx 900$. So, the use of boxes is necessary.

Depending on the interaction and the system, the way that interactions should be put into boxes that yield the best performance can be different \cite{michel2019}. A simple and effective way to group interactions for 2D dipolar systems is to put the close neighbors into a box \cite{komatsu2018, tomita2016}; see Fig.~\spacedsubfigref{fig:updaterulebox}{right}.

To see how boxing near neighbors will benefit these methods, we first discuss its application in the clock method.
Due to the energy compensation, the failure probability for the near-neighbors box is much lower than considering these neighbors individually. For distant neighbors, we can approximate the failure probability $h_j$  by
\begin{equation}
h_j = 1 - e^{-\beta \Delta E_j} \sim \beta r_{ij}^{-3}  \quad (\beta r_{ij}^{-3} \ll 1).
\end{equation}
Here we used the fact that the dipolar interaction decreases with distance as $r^{-3}$. 
The average number of failed events ${N}_f$ outside the box of radius $r_\mathrm{box}$ can be roughly evaluated as
\begin{equation}
{N}_f \sim \int_{r_\mathrm{box}}^{L}  h_j \; r \mathrm{d}r \sim \beta (r_\mathrm{box}^{-1} - L^{-1}).
\end{equation}  
For 2D dipolar systems, the average number of failed events ${N}_f$ in distant neighbors significantly decreases by introducing the near-neighbors box with a moderate radius.
Therefore, boxing only the near neighbors is enough to reduce the number of failed events in both near and distant neighbors.

Introducing the near-neighbors box increases the number of pairwise interaction evaluations, which may appear to result in a longer computational time. But in the clock method, we need to stochastically select the neighbors, mostly from the near neighbors, and verify the bound trials. It requires generations of lots of random numbers which are costly operations. By introducing the near-neighbors box, the number of bound trials significantly decreases. Since the calculation of pairwise energy in the near-neighbors box has a lower computational cost, using a moderately sized near-neighbors box can reduce the computational time while increasing the acceptance rate. To see theoretically how the introduction of the near-neighbors box can reduce the computational time refer to Appendix~\ref{app:theory}.

By putting close neighbors into a box, The factorized Metropolis filter $P$ for the clock method will be written as
\begin{equation}\label{eq:pfacNearFar}
P = P_{\text{box}} \times P_{\text{far}},
\end{equation}
where the first term is defined as
\begin{equation}
\begin{aligned}
\label{eq:pnear}
P_{\text{box}} &= e^{-\beta [\Delta E_{\text{box}}]^+}, \\
\Delta E_{\text{box}} &= \sum _{j \in \mathrm{box}} \Delta E_j,
\end{aligned}
\end{equation}
and
\begin{equation}\label{eq:pfar}
P_{\text{far}} = \prod_{j \notin \text{box}} P_j.
\end{equation}
In other words, we have a large box containing near neighbors and a large number of single-member boxes of distant neighbors. 

Similarly to the clock method, it can be shown that the majority of stochastically selected spins in the SCO and Tomita methods are from near neighbors. Therefore, introducing the near-neighbors box eliminates the need to generate random numbers for stochastic selection of these nearby spins as they are always selected. In addition, the number of selected neighbors increases which results in the increase of the acceptance rate by compensation of the energy difference.

Overall for all the clock, SCO, and Tomita methods, employing the near-neighbors box reduces the effective computational time by both increasing the acceptance rate and decreasing the computational time.
\subsection{Implementation of the effective field method and overrelaxation moves using the box}
As explained in Sec. \ref{sec:Metropolis}, for the Metropolis algorithm, the computational cost of a rejected move can be reduced using the effective field method. The near-neighbors box enables a similar approach for the clock, SCO, and Tomita algorithms. In this case, instead of calculating the interaction of spin $\vec{S}_i$ with the near neighbors individually, we consider the interaction with their effective field  which is defined as
\begin{equation}
\label{eq:effectiveFieldNear}
\vec{H}_i^{\text{box}} = \sum_{j \in \text{box}} \vec{D}_{ij} \cdot \vec{S}_j.
\end{equation}
The summation is performed over the group of near neighbors, represented schematically by the spins inside the dashed circle in Fig.~\spacedsubfigref{fig:updaterulebox}{right}. The effective fields are only updated if a move is accepted, with a procedure similar to Eq. \eqref{eq:updateeffectivefield} for the Metropolis algorithm. As a result, the computational cost of rejected moves are reduced.

The near-neighbors box also enables the implementation of overrelaxation moves for the clock, SCO, and Tomita algorithms. To achieve this, it is sufficient to reflect the spin $\vec{S}_i$  around the effective field generated by nearby spins:
\begin{equation} \label{eq:ORclock}
\vec{S}'_i = -\vec{S}_i + 2 \left( \frac{\vec{S}_i \cdot \vec{H}_i^{\text{box}}}{(H_i^{\text{box}})^2} \right)  \vec{H}_i^{\text{box}}.
\end{equation}
It is notable that in the SCO method, an overrelaxation move can be used even without the box; see Appendix~\ref{app:SCO} for details.

When the spin $\vec{S}_i$ is reflected around the effective field $\vec{H}_i^{\text{box}}$, the interaction energy with its neighbors in the box remains constant (\textit{i.e.}, $\Delta E{_\text{box}} = 0$). However, the interaction energy difference with the stochastically selected far neighbors is a non-zero value.
In the clock method, this implies that $P_\text{box}=1$ and the acceptance probability becomes $P = P_\text{far}$. Therefore, the overrelaxation moves in the clock, SCO, and Tomita algorithms are not rejection-free and exhibit a theoretical dynamical exponent of $z \simeq 2$. However, in Sec. \ref{sec:res:OR}, we demonstrate that it leads to an approximately two-fold reduction in the effective computational time, even though it does not lower the dynamical exponent.
\section{\label{sec:results}Results and Discussion}
In this work, we investigated the triangular lattice of pure dipoles ($J=0$) near the critical point as an example to compare the clock, SCO, and Tomita methods. Some studies have investigated the critical behavior of triangular and square lattices of pure dipoles using the Metropolis method with system lengths up to $L=128$  \cite{tomita2009, rastelli2002,fernandez2007,debell1997,baek2011}. However, a comprehensive understanding of the critical behavior necessitates simulations of larger systems. Yet, to the best of our knowledge, no studies have gone beyond a length of $L>128$. Consequently, our understanding of the critical behavior of these systems remains limited.

We determined the critical point of the triangular lattice of pure dipoles by analyzing the Binder cumulant and logarithmic derivatives of magnetization \cite{newman1999, chen1993} (details provided in Appendix~\ref{app:Oresults}). Our results indicate the existence of a critical point at $T_\mathrm{c}=0.849(1)$, which aligns closely with the findings of Rastelli \textit{et al.} \cite{rastelli2002}. We verified that the results from the clock, SCO, Tomita, and Metropolis methods in the estimated critical temperature were in agreement within the margins of error. Other details of the simulations are given in Appendix~\ref{app:SimDetail}

\begin{figure}[tbp]
    \centering
	\includegraphics[width=\columnwidth]{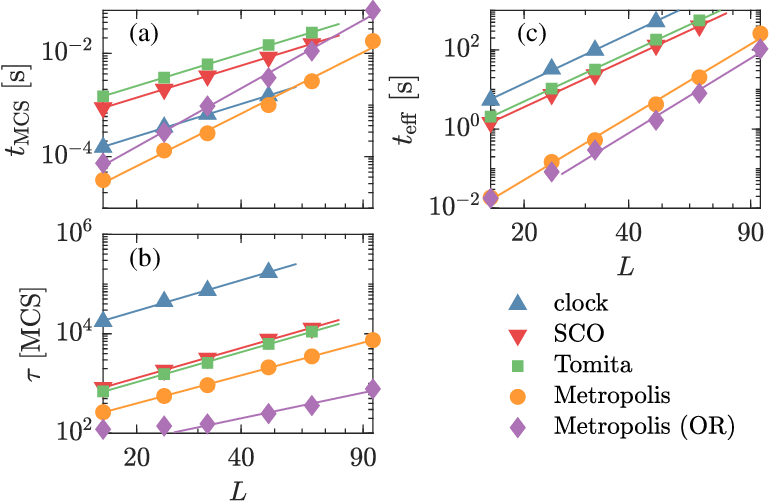}
    \caption{Computation time $t_{\mathrm{MCS}}$ (a), autocorrelation time $\tau$ (b),  and effective computation time $t_{\mathrm{eff}}$ (c) versus system size $L$  for the clock, SCO, Tomita, and Metropolis methods. The near-neighbors box or overrelaxation moves are not used. The Metropolis method with a 10 to 1 overrelaxation to random moves ratio [Metropolis (OR)] is included as the reference of comparison. The simulations are taken at the estimated critical point of the triangular lattice of dipoles at $T_\mathrm{c}=0.849$. The error bars are ignored since they are smaller than the size of the symbols, and the lines represent best fits to the data.}
    \label{fig:originalCompare}
\end{figure}

In Fig.~\ref{fig:originalCompare}, we compared the Metropolis, clock, SCO, and Tomita methods without the boxing technique or overrelaxation moves.  However, the Metropolis method with a 10:1 overrelaxation to random moves ratio was used as the reference method and is denoted as ``Metropolis (OR)" in the figure. As shown in Fig.~\ref{fig:originalCompare}, the computational time for a single MC step of the clock method $t_\mathrm{MCS}$  is about an order of magnitude smaller than those of the SCO and Tomita methods. However, its autocorrelation time $\tau$  is also about an order of magnitude larger. Consequently, the effective computational times $t_\mathrm{eff}$ of the clock, SCO, and Tomita methods are roughly the same order of magnitude.

The effective computational time of the Metropolis (OR) method is significantly lower than the clock, SCO, and Tomita methods without the boxing technique and overrelaxation moves, as evident in Fig.~\subfigref{fig:originalCompare}{c}. However, in the log-log plot, the effective computational time of the Metropolis method has a steeper slope compared to the other three methods. This suggests that for large systems, the Metropolis (OR) method will be less efficient. Yet, by extrapolating the effective computational time, we found that the Metropolis (OR) method is only less efficient for system lengths $L \geq 900$. Note that simulating systems with $L  \geq  900$ is impractical as they require very powerful computing systems that are not readily available. Therefore, without the boxing technique and overrelaxation moves, the clock, SCO, and Tomita methods are not suitable for investigating the critical behavior of the pure dipolar triangular lattice.

\begin{figure}[tbp]
    \centering
	\includegraphics[width=\columnwidth]{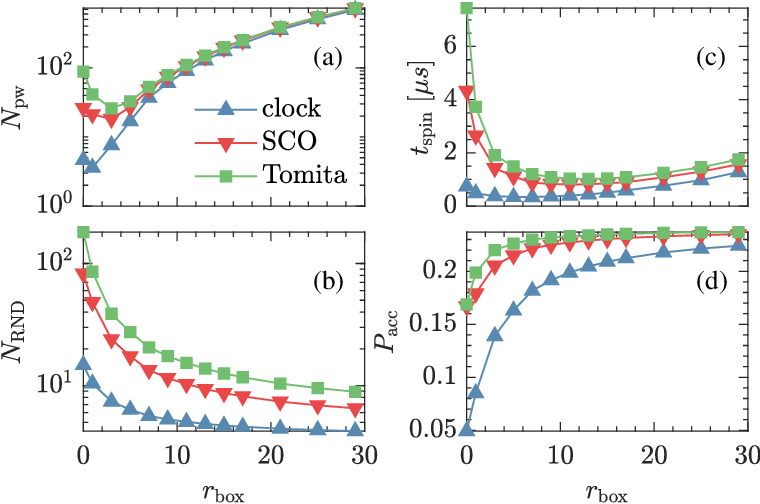}
    \caption{(a) Average number of pairwise interaction evaluations per spin $N_\mathrm{pw}$, (b) average random number generations per spin $N_\mathrm{RND}$, (c) computational time per spin $t_\mathrm{spin}$, and (d) acceptance ratio $P_\mathrm{acc}$ plotted against the near-neighbors box radius $r_\mathrm{box}$ for the clock, SCO and Tomita methods using random moves only. The Fukui-Todo method is employed for the stochastic selection of far spins. The simulations are performed at the system size of $L=256$ and the estimated critical temperature, $T_\mathrm{c}=0.849$, for the triangular lattice of pure dipoles. The legend is the same for all the curves. The errors are smaller than the size of the symbols, and the lines are only drawn to guide the eye.}
    \label{fig:BoxCompare}
\end{figure}

\subsection{Employing the near-neighbors box}
We can apply the near-neighbors box to reduce the computational time and increase the acceptance rate of the clock, SCO, and Tomita methods. Both effects lead to the reduction of effective computational time.  We consider the near-neighbors box to be a circular area of radius $r_\mathrm{box}$ with the changing spin at its center. In Fig.~\ref{fig:BoxCompare}, the radius of the near-neighbors box $r_\mathrm{box}$ is varied and its effect on the average computational time of a spin update $t_\mathrm{spin}$ and other related quantities are shown for system length $L=256$. The two major quantities affecting $t_\mathrm{spin}$ which we investigated are the average number of pairwise interaction evaluations per spin, $N_\mathrm{pw}$, and the average random number generations per spin, $N_\mathrm{RND}$.

In Fig.~\subfigref{fig:BoxCompare}{a}, it can be seen that $N_\mathrm{pw}$ increases with increasing box radius ($r_\mathrm{box}$). The initial dip in the curves is due to the effective field method, which reduces the cost of rejected moves. Without the box, nearest neighbors are almost always selected, and their pairwise interactions are evaluated whether in an accepted or rejected move. However, when placed in the box, their pairwise interactions are evaluated only in accepted moves because of the effective field method. This initially reduces $N_\mathrm{pw}$, but when $r_\mathrm{box}$ increases further, eventually the box becomes too large, and in each acceptance, the pairwise interactions for a large number of spins must be calculated, so $N_\mathrm{pw}$ increases. 

Fig.~\subfigref{fig:BoxCompare}{b} shows the average random number generations per spin ($N_\mathrm{RND}$). The $N_\mathrm{RND}$ for SCO and Tomita methods without the box is much larger than for the clock method. This is because, in the SCO and Tomita methods, all of the stochastically selected neighbors must be found before deciding to accept a move. However, for the clock method, the process can end in a failure event, avoiding the need to find all of the stochastically selected neighbors. This is also the reason why the $N_\mathrm{pw}$ is smaller in the clock method in Fig.~\subfigref{fig:BoxCompare}{a}.

By enlarging the box of nearby neighbors, the number of stochastically selected distant spins is greatly reduced, reducing $N_\mathrm{RND}$ as can be seen in Fig.~\subfigref{fig:BoxCompare}{b}. Because random number generations are costly operations, this reduction in $N_\mathrm{RND}$ has a significant effect on reducing the average computational time of the spin update, $t_\mathrm{spin}$, especially for the SCO and Tomita methods, as shown in Fig.~\subfigref{fig:BoxCompare}{c}. Although	$t_\mathrm{spin}$ is also reduced for the clock method, its reduction is not as significant as the SCO and Tomita methods, since $N_\mathrm{RND}$ for the clock method is small even without the box.

The average computational time of a spin update $t_\mathrm{spin}$ is computed by dividing the computational time of the simulation by the number of spin updates in the simulation. The plots of $t_\mathrm{spin}$ \textit{vs.} $r_\mathrm{box}$ are presented in Fig.~\subfigref{fig:BoxCompare}{c}. The curves exhibit a minimum, which is also expected theoretically (see Fig.~\ref{fig:r2plus1r} in Appendix~\ref{app:theory}).

As can be seen from Fig.~\subfigref{fig:BoxCompare}{d}, the acceptance rate for all three methods is increased by introducing the near-neighbors box. The increase in the acceptance rate is due to the compensation of the energy difference, see Eq. \eqref{eq:MetropolisFacinequality}. Unlike the clock method, there is some compensation in the SCO and Tomita methods even without the near-neighbors box, which results in a higher acceptance rate for these methods.

In general, the introduction of the nearby-neighbors box proves advantageous for the clock, SCO, and Tomita methods by reducing computational times and increasing acceptance rates. Specifically, the use of the box notably enhances the acceptance rate in the clock method, while in the SCO and Tomita methods, it primarily leads to the reduction of $t_\text{spin}$.

Suitable box sizes can be found where both the acceptance rate is high and the computational time is short. As shown in Figs.~\subfigref{fig:BoxCompare}{c} and \subfigref{fig:BoxCompare}{d}, the range of $r_\mathrm{box} = 11$ to $r_\mathrm{box} = 17$ offers short computational times ($t_\mathrm{spin}$ ) for all methods while maintaining a relatively high acceptance rate.

\begin{figure}[tbp]
    \centering
    \includegraphics[width=\columnwidth]{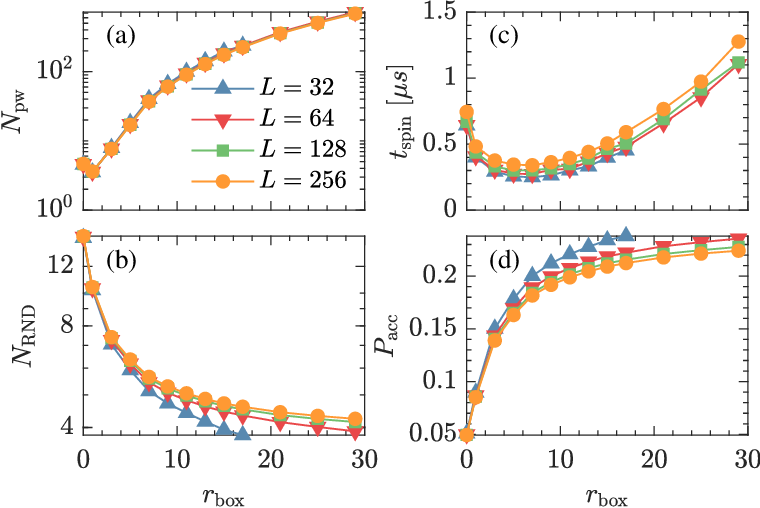}
    \caption{Parameters involved in algorithm efficiency as a function of box radius $r_\mathrm{box}$ for the clock method in different system sizes. The parameters and other details are the same as Fig.~\ref{fig:BoxCompare}. Errors are small compared to the size of the symbols.}
    \label{fig:BoxVsL}
\end{figure}

For the clock method, the curves of $t_\mathrm{spin}$ and related quantities in terms of $r_\mathrm{box}$ are shown for different system sizes in Fig.~\ref{fig:BoxVsL}. 
Considering a fixed-size near-neighbor box, it can be observed that the variables exhibit negligible dependence on system size, which is expected from the theory \cite{michel2019}, and can also be seen in Appendix~\ref{app:theory}.
Therefore, in 2D dipolar systems, by finding a suitable $r_\mathrm{box}$ for an arbitrary system length, this $r_\mathrm{box}$ can also be used for other system sizes. The independence of these variables with system size is also observed in the SCO and Tomita methods (data not shown).

\subsection{Adding Overrelaxation moves} \label{sec:res:OR}
\begin{figure}[tbp]
    \centering
	\includegraphics[width=\columnwidth]{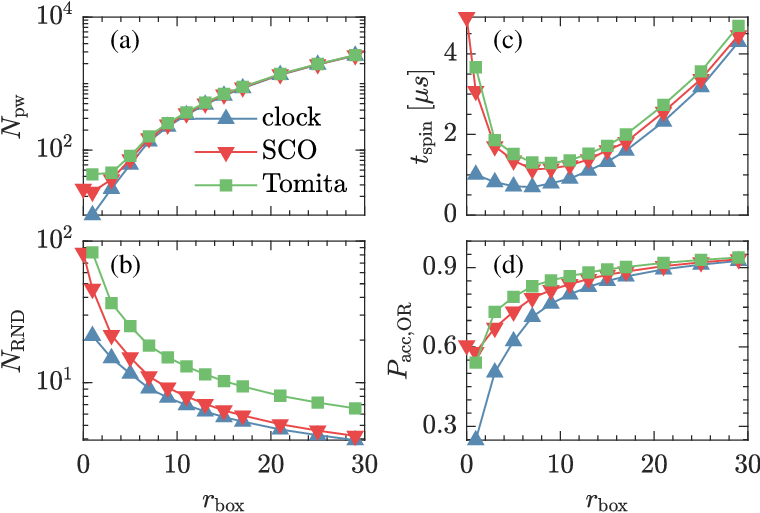}
    \caption{Parameters involved in algorithm efficiency \textit{versus} box radius $r_\mathrm{box}$ for the clock, SCO and Tomita methods with overrelaxation moves. The ratio of overrelaxation to random moves is set to 10. The acceptance ratio of the overrelaxation moves is indicated by $P_\mathrm{acc,OR}$. Errors are smaller than the data point symbols. Other details and the parameters are the same as in Fig.~\ref{fig:BoxCompare}.}
    \label{fig:BoxOR}
\end{figure}

Next, we investigated the effect of adding overrelaxation moves to the clock, SCO, and Tomita methods.
Fig.~\ref{fig:BoxOR} illustrates this effect for various near-neighbors box radii. For the clock and Tomita methods, overrelaxation moves can only be added when using the near-neighbors box. But for the SCO method, it can be added even without the box. In these simulations, we set the ratio of overrelaxation and random moves to 10:1.

The results of adding overrelaxation moves for different box radii in Fig.~\ref{fig:BoxOR} are generally similar to using random moves only, shown in Fig.~\ref{fig:BoxCompare}. However, overrelaxation moves have a higher computational cost due to their higher acceptance rate. This is because, with the effective field method, the cost of rejected moves is significantly reduced compared to that of accepted ones.  Therefore, the values of $N_\mathrm{pw}$,  $N_\mathrm{RND}$ and  $t_\mathrm{spin}$ for overrelaxation moves in Fig.~\ref{fig:BoxOR} are higher than random moves in Fig.~\ref{fig:BoxCompare}. Additionally, the curves of $t_\mathrm{spin}$ \textit{versus} $r_\mathrm{box}$  are less flat for overrelaxation moves.

The acceptance rate of the overrelaxation moves is depicted in Fig.~\subfigref{fig:BoxOR}{d}. As shown, the SCO and Tomita methods have a higher overrelaxation acceptance rate than the clock method.  The acceptance rate of the random moves in these simulations is not shown here, as it was very similar to that in Fig.~\subfigref{fig:BoxCompare}{d}. The suitable box sizes in this case are found for values of $r_\mathrm{box}$ between 7 and 11, where all the methods achieve both a short computational time and a high acceptance rate.

\begin{figure}[tbp]
    \raggedright
	\includegraphics[width=\columnwidth]{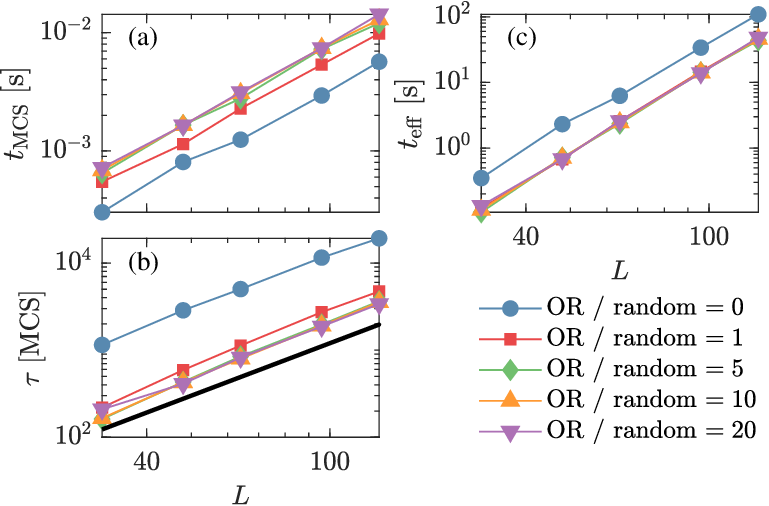}
    \caption{(a) Computation time of an MC step, (b) autocorrelation time, and (c) effective computational time \textit{versus} system length for the clock method using different ratios of overrelaxation (OR) to random moves. These ratios include 0, 1, 5, 10, and 20. The ratio 0 corresponds to using only random moves. The lines are only a guide to the human eye. Specifically, the solid line in (b) has a slope of 2 in the log-log plot. The box radius is set to $r_{\mathrm{box}}=11$, and the temperature is set to $T_\mathrm{c} = 0.849$. Other details are similar to Fig.~\ref{fig:originalCompare}.}
    \label{fig:ORRatios}
\end{figure}

\begin{figure}[tbp]
    \centering
	\includegraphics[width=\linewidth]{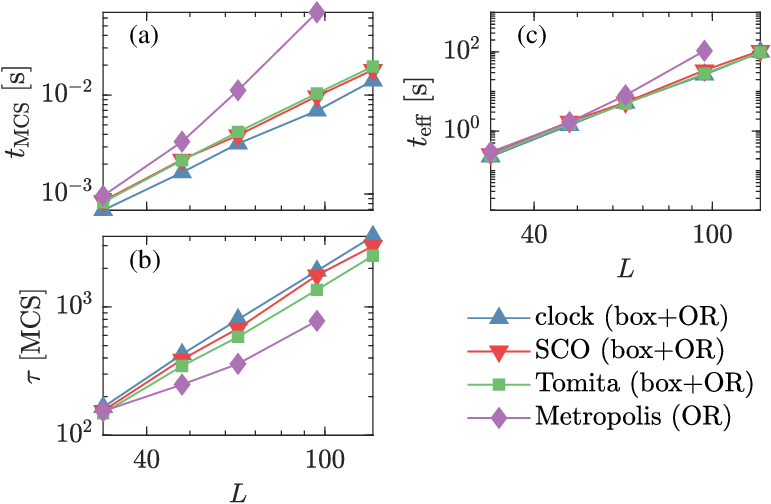} 
    \caption{(a) Computation time $t_{\mathrm{MCS}}$, (b) autocorrelation time $\tau$, and (c) effective computation time $t_{\mathrm{eff}}$ against system size $L$ for the clock, SCO, Tomita, and Metropolis methods. The box radius is set to $r_\mathrm{box}=11$, and the ratio of overrelaxation to random moves is set to 10. The simulations are taken at the estimated critical point of the triangular lattice of dipoles at $T_\mathrm{c}=0.849$. Legends are the same for all the sub-figures. The lines serve as a visual guide only and the error bars are small compared to the size of data points.}
    \label{fig:CompareFinal}
\end{figure}

\begin{figure}[tbp]
    \centering
\includegraphics[width=\linewidth]{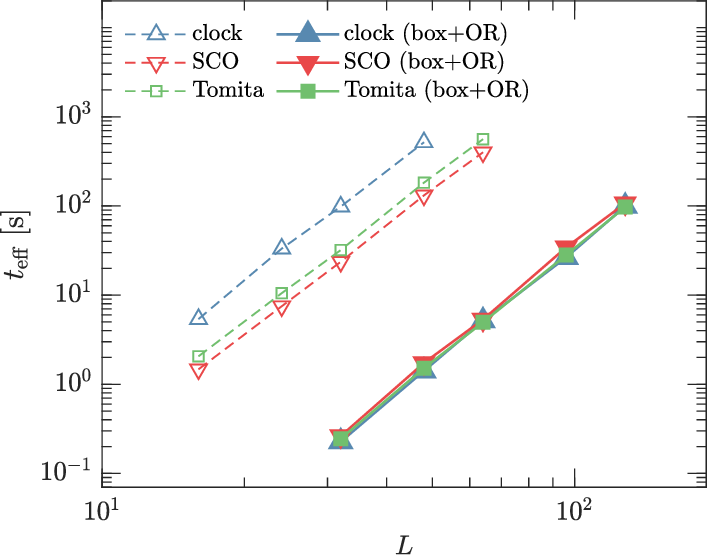} 
    \caption{The effective computational time of the clock, SCO, and Tomita methods for different system lengths. The results with the near-neighbors box ($r_\text{box}=11$) and overrelaxation moves (10:1 ratio) are shown with the solid symbols, while the results of the simple implementation of methods (without the box and overrelaxation) are presented with the empty symbols. Other details are similar to Fig.~\ref{fig:CompareFinal}}
    \label{fig:FinalVsPlain}
\end{figure}

Next, we investigated the influence of different ratios of overrelaxation to random moves on the performance of the clock method (Fig.~\ref{fig:ORRatios}). In these simulations, we set the box radius to $r_\text{box} = 11$. Adding overrelaxation moves decreases the autocorrelation time, as shown in Fig.~\subfigref{fig:ORRatios}{b}. However, since these moves are not rejection-free, the dynamical exponent $z$ remains close to $z \approx 2$; the solid line with a slope of 2 in the figure, clearly illustrates this point.  So, unlike the Metropolis method, where the overrelaxation also reduces the dynamical exponent, here the addition of overrelaxation moves only leads to a constant reduction in the multiplicative factor of the autocorrelation time. Additionally, Fig.~\ref{fig:ORRatios} shows that the computational time increases for larger ratios because overrelaxation moves are more expensive than random moves. Overall, the effective computational time becomes nearly equal for the different ratios, as the increase in computational time is balanced by the decrease in autocorrelation time.

In summary, adding overrelaxation moves to the clock method reduces the effective computational time by nearly half for all the overrelaxation ratios. We observed similar results with the SCO and Tomita methods.

\subsection{Comparison of the enhanced and simple implementations of the three methods}
We have demonstrated that employing the near-neighbors box and adding overrelaxation moves reduces the effective computational time of the clock, SCO, and Tomita methods. Therefore, we simulated these methods with a box radius of $r_\text{box} = 11$ and overrelaxation moves and compared them to the Metropolis method with overrelaxation. The results of this set of simulations are shown in Fig.~\ref{fig:CompareFinal}. From Fig.~\subfigref{fig:CompareFinal}{c}, it can be observed that for all the simulated sizes, the clock, SCO, and Tomita methods were as efficient or more efficient than the Metropolis method. The efficiency advantage of the clock, SCO, and Tomita methods over the Metropolis method becomes more significant at larger sizes. Therefore, simulating larger systems to investigate critical behaviors is less computationally expensive with these methods. 

With the use of the near-neighbors box and overrelaxation moves, the efficiency of the clock, SCO, and Tomita methods is significantly improved. In comparison to their untuned implementations, the efficiency of the clock method has improved by a factor of approximately 350, the SCO method has seen a 75-fold speedup, and the Tomita method has become 100 times faster.  Fig.~\ref{fig:FinalVsPlain} clearly illustrates these significant performance gains.
 
\subsection{Comparison of the clock, SCO, and Tomita methods}
The clock, SCO, and Tomita methods are all claimed to have a computational time complexity of $\mathcal{O}(N)$ or $\mathcal{O}(L^2)$ for 2D dipolar systems. Table \ref{tab:scaling} clearly demonstrates this by presenting the scaling exponents of computational time, autocorrelation time, and effective computational time for these algorithms. It can also be observed that unlike the Metropolis algorithm, where the dynamical exponent decreases from 2 using the overrelaxation moves, for the clock, SCO, and Tomita methods, adding overrelaxation moves keeps the scaling of autocorrelation time at approximately 2. 

It is important to note that the scalings in Table \ref{tab:scaling} are obtained from Figs. \ref{fig:originalCompare} and \ref{fig:CompareFinal}, where the fitting range for system length $L$ is less than one order of magnitude. Moreover, for the Metropolis algorithm (with and without overrelaxation), the best fitting is achieved with an even smaller fitting range of  $L \sim 32-96$. Therefore, these scalings should be considered approximate. Considering the errors and approximations, the scaling exponents for the clock, SCO, and Tomita methods, both in their simple implementations and with the box and overrelaxation moves, are equal.

We observe that the efficiency of the clock, SCO, and Tomita methods, all employing the near-neighbors box and overrelaxation moves, are very similar (see Fig.~\ref{fig:FinalVsPlain}). This similarity highlights that the major advantage of one algorithm over another lies in the simplicity of its concept and implementation. In this regard, when comparing the theory behind the three methods, the clock and SCO methods are simpler. The Tomita method is conceptually more complex because it was developed for the long-range Ising model and extended for use in dipolar systems.

In the case of implementation, all the clock, SCO, and Tomita methods rely on the dynamic thinning or Fukui-Todo techniques for the stochastic selection of spins. Since implementing these two techniques is the most complex step, the clock, SCO, and Tomita methods have nearly similar implementation complexity. The code for all the methods compared in this article is written in C++ and is available on GitHub \cite{onmc}.

While this article describes the application of the clock, SCO, and Tomita methods for fixed spin systems in lattices, these methods can potentially be extended to liquid systems by introducing a grid and considering the particles to be located within the grid cells \cite{michel2019, sasaki2008}. However, it is important to note that the authors have not applied these methods to liquids, and to the best of our knowledge, there is no existing research on their application to such systems.

For the implementation of the clock, SCO, and Tomita methods in liquid systems, the ECMC method can be used as a base \cite{bernard2009}. Similar to the clock method, this method is based on the factorized Metropolis filter and utilizes thinning methods to reduce complexity \cite{michel2014, kapfer2016}. Although the ECMC method is not directly suitable for rotating dipoles \cite{hollmer2022}, it has been tested for liquid systems and has publicly available codes \cite{kapfer2016, hollmer2020}. The codes for the ECMC method alongside our code for lattices might accelerate the development of the clock, SCO, and Tomita methods for liquids.

\begin{table} % table 2
\centering
\caption{Scaling exponents of computational time ($t_{\text{MCS}}$), autocorrelation time ($\tau$), and effective computational time ($t_{\text{eff}}$) in the different algorithms for the dipolar triangular lattice. The term ``box" denotes using the near-neighbors box, and ``OR" denotes the addition of overrelaxation moves with a 10 to 1 ratio. The scalings are obtained from Figs. \ref{fig:originalCompare} and \ref{fig:CompareFinal}, but they are rather approximate since the fitting range for $L$ is limited to less than one order of magnitude.}
\begin{tabularx}{\columnwidth}{p{3cm}XXX}
\toprule
		      	  & \multicolumn{3}{l}{\qquad \quad Scaling exponents} \\ 
Algorithm     & ~\:\!$t_{\text{MCS}}$ & ~~\;$\tau$ & ~\,$t_{\text{eff}}$ \\ \midrule
clock         &2.0(1) & 2.0(1) & 4.0(2)  \\
SCO     &2.0(1) & 2.0(1) & 4.0(2)  \\
Tomita    &2.0(1) & 2.0(1) & 4.0(2)  \\
Metropolis  &3.7(3)  & 1.9(1) & 5.7(4)  \\ 
clock (box+OR)      &2.1(1) & 2.2(2) & 4.3(3)  \\
SCO (box+OR)      &2.2(1) & 2.1(1) & 4.3(2)  \\
Tomita (box+OR)      &2.3(2) & 2.0(1) & 4.3(3)  \\
Metropolis (OR) &3.9(2)  & 1.4(2) & 5.4(4)  \\ \bottomrule
\end{tabularx}
\label{tab:scaling}
\end{table}

\section {conclusion}
We examined the effectiveness of various Monte Carlo methods ---Metropolis, clock, SCO, and Tomita--- in simulating a system of pure dipoles arranged on a triangular lattice near the critical point. 

The clock, SCO, and Tomita methods demonstrated a computational complexity of $\mathcal{O}(N)$  or $\mathcal{O}(L^2)$, whereas the Metropolis method exhibited $\mathcal{O}(N^2)$  complexity for this system. Our findings revealed that, despite the lower computational complexity, the untuned versions of the clock, SCO, and Tomita methods were less efficient than the Metropolis method with overrelaxation moves. This inefficiency was primarily due to the lower acceptance rate in the clock method and the longer computational times in the SCO and Tomita methods compared to the Metropolis algorithm for the dipolar triangular lattice.

We investigated the effect of employing the boxing technique to group nearby neighbors in the clock, SCO, and Tomita methods. Our analysis indicated that the use of the near-neighbors box led to a reduction in computational time and an increase in the acceptance rate for these methods. This combined effect resulted in a significant improvement in the efficiency of these three methods.

Furthermore, we explored adding overrelaxation moves to the clock, SCO, and Tomita methods with the near-neighbors box. Our results showed that while this addition increased the computational cost, it also decreased the autocorrelation time by a constant factor.  This resulted in a nearly twofold efficiency improvement.

We found that the efficiency of these three methods, which employ the near-neighbors box and overrelaxation moves, has significantly improved compared to their untuned versions. Specifically, the clock method has become approximately 350 times faster, the SCO method 75 times faster, and the Tomita method 100 times faster. We found that these enhanced versions surpass the Metropolis algorithm (with overrelaxation moves) for the simulation of dipolar triangular lattice near the critical point.

The three methods in their untuned implementations, exhibited distinct performance for the simulation of dipolar triangular lattice, with SCO being the most efficient and clock method having the lowest performance. However, when enhanced with the near-neighbors box and overrelaxation moves, all three methods achieved nearly identical efficiency. Due to this equivalence, the selection of an algorithm depends on its theoretical concept and implementation simplicity. The clock and SCO methods are easier to understand, while the implementation complexity of all three methods is roughly equivalent. This is because they all rely on  the dynamic thinning or Fukui-Todo techniques which are the core of the algorithm. The implementations of all the algorithms in this work can be found in our public code \cite{onmc}.

In summary, while the untuned forms of the clock, SCO, and Tomita methods are less efficient than the Metropolis method for studying the 2D dipolar triangular lattice near the critical point, employing the near-neighbors box method with overrelaxation moves significantly reduces their autocorrelation time and computational time, making them competitive tools for investigating critical behavior of 2D dipolar systems. Their enhanced versions offer similar efficiency, making the choice between them a matter of theoretical clarity and implementation ease (clock and SCO methods).

\begin{acknowledgments}
This research received support from the IASBS Grant G2022IASBS12644 and was facilitated by the resources of Sharif HPC. The draft manuscript was improved to a more fluent form by utilizing AI tools, namely ChatGPT, Microsoft copilot, and Google Gemini to identify grammatical errors, rephrase sentences, and enhance the overall quality of the text.
The authors would like to express their gratitude to Dr. Bahman Farnudi for his assistance in editing this paper. We would also like to thank Professor Y. Tomita for providing valuable information that helped us understand his method.
 
\end{acknowledgments}

\appendix
\section{Other results \label{app:Oresults}}
For detection of the critical point, we used the Binder cumulant which is defined as
\begin{equation}\label{eq:Binder}
U_4 = 1 - \frac{\langle M^4 \rangle}{3 \langle M^2 \rangle},
\end{equation}
where $\langle M \rangle= \left\langle |\sum_i \vec{S}_i| \right\rangle $ is the magnetization of the system. The Binder cumulant reaches constant values at the limits of $T=0$ and $T=\infty$. The curve of this parameter for each system size has a unique slope. From the finite size scaling hypothesis, it is known that the Binder cumulant for different system sizes has an equal value at the critical point. Therefore, the temperature where these curves intersect indicates the critical point \cite{binder2010, binder1981}.

\begin{figure}[tbp]
	\centering
\includegraphics[width=\columnwidth]{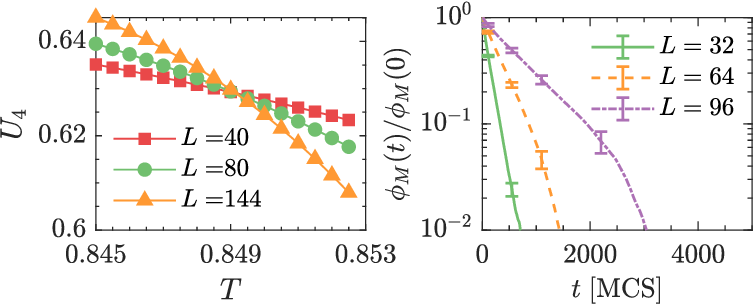}
	\caption{\label{fig:Binder} (a) Binder parameter $U_4$ \textit{vs.} temperature for different system sizes, and (b) normalized autocorrelation function of magnetization as a function of time for different system sizes at the critical point. The errors in (a) are smaller than the symbol sizes, and lines are included to guide the eye. The simulations in (b) were performed using the Metropolis algorithm with overrelaxation, and error bars are plotted only for a select number of data points for clarity.} \label{fig:BinderAutocorrelation}
\end{figure}

For the triangular lattice of dipoles, this crossing happens at the temperature $T_\mathrm{c}=0.849(1)$, which can be seen in Fig.~\subfigref{fig:BinderAutocorrelation}{a}. We further confirmed the existence of this critical point with more advanced methods, including the logarithmic derivatives of magnetization \cite{chen1993} and the short-time critical dynamics method \cite{zheng1998}.

The normalized autocorrelation function of magnetization $\phi_M(t) / \phi_M(0)$ at the estimated critical temperature $T_\mathrm{c}=0.849$  in the Metropolis algorithm (with overrelaxation moves) is depicted in Fig.~\subfigref{fig:BinderAutocorrelation}{b} for some system sizes. As the figure shows, the slope of the function in semi-log plot decreases with increasing system size. This indicates that the magnetization takes longer to lose correlation with its past values in larger systems.

\section{Simulation details \label{app:SimDetail}}
In the simulation of the triangular lattice of pure dipoles, the shape of the simulation cell is chosen to be a parallelogram (see Fig.~\ref{fig:updaterulebox}), and the periodic boundary condition is applied. This parallelogram is defined by basis vectors $\vec{a} = L \vechat{x} + L/2 \, \vechat{y}$ and $\vec{b} = L \sqrt{3}/2 \, \vechat{y}$. In the ground state of the dipolar triangular lattice with the periodic boundary condition, all spins align within the lattice plane \cite{rastelli2002, politi2006}. Therefore the order parameter is calculated as $\langle M \rangle= \left\langle |\sum_i \vec{S}_i| \right\rangle $. We performed simulations using a reduced unit system, where we set  $\mu_0/4\pi=1$ and the lattice constant to unity. The spins are chosen to be plane rotators, which means that they are constrained to rotate within the plane of the lattice. To calculate the dipolar dyadic tensors $D_{ij}$ in periodic boundaries, we applied the Ewald summation technique \cite{weis2003, debell2000, mol2014}. 

To verify the algorithm implementations for high-accuracy simulations, we performed a set of simulations on the $16\times16$  lattice at $T = 1.4$  for the clock, SCO, Tomita, and Metropolis methods. Each simulation used 50 runs with $10^7$ MC steps for averaging and employed a large box size (for the clock, SCO, and Tomita methods). These simulations were conducted both with and without the use of overrelaxation moves. We found that the energy and magnetization results obtained from the different algorithms were in agreement up to 4 significant digits.

Our analysis demonstrates that the Fukui-Todo approach for stochastic selection of spins is more efficient for larger box sizes in comparison with the dynamic thinning method. Consequently, all simulations presented in Sec. \ref{sec:results} utilize the Fukui-Todo method. For a detailed comparison of the dynamic thinning and Fukui-Todo techniques, refer to Appendix~\ref{app:FTvsDythin}.

The autocorrelation time of the algorithms was determined  using $\tau=\max(\tau_E, \tau_M)$, where $\tau_E$ and $\tau_M$ represent the autocorrelation times of energy and magnetization, respectively. To compute the autocorrelation time for each algorithm, we simulated 8 runs, each containing between $3\times10^5$ to $10^6$ MC steps for equilibration and $3\times10^6$ to $10^7$ MC steps for averaging. Since the clock method without the box exhibits a higher autocorrelation time, we used up to $10^7$ MC steps for equilibration and $10^8$ MC steps	 for averaging to get accurate results.

In order to measure the computational time, we executed simulations with a single thread on a commercial computer equipped with an Intel i7 6700K CPU and 24 GB of RAM. The simulations had lengths of $10^3$ to $2\times 10^5$ MC steps depending on the algorithm and system size.

The energy of the system in the clock, SCO, and Tomita methods is obtained using the FFT method \cite{krech2000, sasaki2008}, which has a computational complexity of $\mathcal{O}(N \ln(N))$. Because the system sizes we investigated were not very large, the additional complexity introduced by the energy computation was not observed. Therefore, despite the expected $\mathcal{O}(N \ln(N))$ complexity from the FFT method, the scalings of the computational time in Table \ref{tab:scaling} were derived based on an assumption of $\mathcal{O}(N)$ complexity. However, it is important to note that this observation may not hold for much larger systems.

Regarding the clock method, we also investigated the constant boxes strategy suggested in Ref. \cite{michel2019}. Our findings indicated negligible differences between the constant boxes approach and using only the near-neighbors box for this system.

\section{Dynamic thinning method \label{app:dythin}}
In the clock algorithm, the dynamic thinning method is used to quickly decide whether to accept or reject a change according to the factorized Metropolis filter \cite{michel2019, shanthikumar1985}. This method allows us to make this decision without having to directly compute Eq. \eqref{eq:fac}. In this method, a fixed value of $\hat{h}$ is assigned to each spin. If a false failure happens, this value is lowered and we find the next bound failure event. 

We first arrange the maximum failure probabilities $\{\max{(h_k)}\}$ in decreasing order as
\begin{equation}\label{eq:sortedh}
\max{(h_1)} > \max{(h_2)}  > \cdots > \max{(h_{N-1})}.
\end{equation}
This means that we are labeling the neighbors of spin $i$ from the one with the strongest interaction to the one with the weakest interaction.

To determine the fate of a change, we begin by setting an auxiliary variable $j' = 0$ and assigning all bound failure probabilities to $\hat{h} = \max({h}_{j'+1})$. Then we find the first failure event of bound trials by  
\begin{equation}\label{eq:increase_j_thinning}
j =  j' + \left\lceil \frac{\log(1-u)}{\log(1-\hat{h})} \right\rceil.
\end{equation}
Here, the notation $\lceil \phantom{i} \rceil$ indicates the ceiling function and $u$  represents a uniform random number in the interval $[0,1)$. 

\begin{algorithm}[H]
\caption{Calculation of factorized Metropolis filter using the dynamic thinning method} \label{alg:dythin}
\begin{algorithmic}[1]
\newcommand{\CommentGreen}[1]{\Comment{\textcolor[rgb]{0.180, 0.545, 0.341}{#1}}}
\State \textbf{Input}: A proposed update $\vec{S}_i \rightarrow \vec{S}'_i$.
\State \textbf{Output}: Update accepted or rejected by returning True or False.
\State \textbf{Initialization}: Label neighbors of $\vec{S}_i$  based on their maximum failure probabilities as $\max{(h_1)} > \max{(h_2)}  > \cdots > \max{(h_{N-1})}$.
\Statex
\State Generate a uniform random variate $u \in [0, 1)$.
\If{$u < 1 - e^{-\beta \Delta E_{\text{box}}}$} 
	\State \textbf{return} False \CommentGreen{Rejection by near neighbors}
\EndIf
\Statex
\State $j' \gets N_\text{box}$ \CommentGreen {Size of the box}
\While{$j' < N$}
	\State $\hat{h} \gets \max({h}_{j'+1})$ \CommentGreen{Assign bound failure probabilities}
    \State Generate a uniform random variate  $u \in [0, 1)$
    \State $j \gets j' + \left\lceil \dfrac{\log (1-u)}{\log(1-\hat{h})}\right\rceil$ \CommentGreen{Next bound rejection}
    \If{$j \geq N$}
        \State \textbf{break}
    \EndIf
    \State Generate a uniform random variate $u \in [0, 1)$
    \State Evaluate $h_{j} = 1 - P_{j}$
    \If{$u < \left( \dfrac{h_{j}}{\hat{h}} \right) $} \CommentGreen{Check if it is a true rejection}
            \State \textbf{return} False \CommentGreen{Rejection by far neighbors}
    \EndIf
	\State $j' \gets j$
\EndWhile
\State \textbf{return} True \CommentGreen{Update is accepted}
\end{algorithmic}
\end{algorithm}

\begin{figure}[tbp]
	\centering
	\includegraphics[width=\columnwidth]{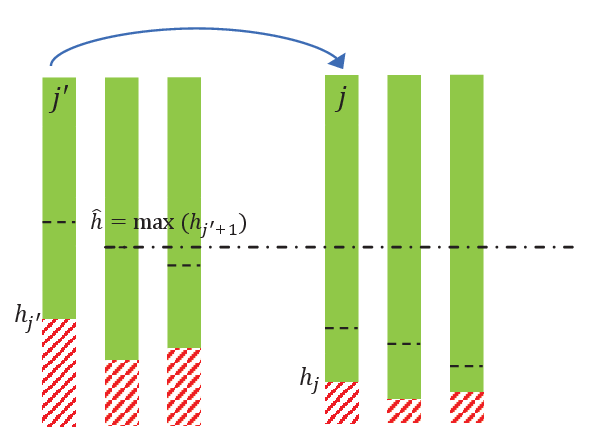}
	\caption{\label{fig:h_jDyn}  Dynamic thinning method. The shaded and hatched regions and the dashed lines are defined similarly to those in Fig.~\ref{fig:clock}.  The dashed-dotted lines represent the maximum failure probabilities. $j'$ is the last false failure event. The next bound failure event $j$ is found by setting all bound failure probabilities to $\hat{h} = \max(h_{j'+1})$ and inverting the PDF for the next bound failure event, which follows a geometric distribution.} 
\end{figure}

Similar to the explanation of Eq. \eqref{eq:geometric_rejection}, if the bound trial is succeeded, it means the result of the actual trial is also a success. Otherwise, we sample a verification trial with the failure probability of 
\begin{equation*}\label{eq:hconThinning}
h_{j,\text{ver}} = \frac{h_{j}}{\hat{h}}
\end{equation*}
to verify if it is a failure in the actual trial or not.
If it is an actual failure, the process is finished and the trial is rejected. 
As depicted in Fig.~\ref{fig:h_jDyn}, we then set $j'=j$ and allocate all bound failure probabilities to $\hat{h} = \max({h}_{j'+1})$. Following this, we identify the next failure event of bound trials $j$ by applying Eq. \eqref{eq:increase_j_thinning}. We repeat this process until $j \geq N$, which indicates the change has been accepted \cite{michel2019}.

For the clock algorithm employing the near-neighbors box, we first calculate  $\Delta E_\text{box}$ as defined in Eq. \eqref{eq:pnear} and check if the result of sampling from $P_\text{box}$ is successful or not. If it is a success, we start the dynamic thinning method from the first far neighbor, \textit{i.e.} we set $j' = N_\text{box}$ as it points to the last member of the box,  and follow a process similar to the clock algorithm. Algorithm~\ref{alg:dythin} presents the calculation of the factorized Metropolis filter using the dynamic thinning method for the clock algorithm with the near-neighbors box.

To perform an overrelaxation move, we simply need to propose an update $\vec{S}_i \rightarrow \vec{S}'_i$ based on Eq. \eqref{eq:ORclock}, instead of a random vector. Since the interaction energy with the nearby neighbors remains unchanged during an overrelaxation move, the trial in line 5 of Algorithm~\ref{alg:dythin} will always be successful.

\section{Walker alias method \label{app:walker}}
The Walker alias method is an efficient algorithm for sampling from a discrete probability distribution with $\mathcal{O}(1)$ time complexity \cite{walker1977, vose1991}. However, it is not directly applicable to finding bound failure events in the clock method. Instead, it forms the foundation for the Fukui-Todo approach \cite{fukui2009} which can be applied to the clock method \cite{michel2019}. The details of the Fukui-Todo method are explained in Appendix~\ref{app:fukui}.

Suppose we have a probability distribution that has $n$ elements. If all the outcomes occur with an equal probability of $1/n$, then sampling from this distribution can be done efficiently with  $\mathcal{O}(1)$ complexity. However, if the probabilities of each outcome differ, then sampling in $\mathcal{O}(1)$ is not trivial. The Walker alias method addresses this challenge by converting the problem into sampling from a distribution with equal probabilities, which can again be achieved in $\mathcal{O}(1)$ time.

We denote the probabilities of the $n$ outcomes as $q_i$, where $i = 1, 2, ..., n$. These probabilities are normalized, summing to unity. We can then categorize them into two groups: the `small' group, for which $q_i \leq 1/n$, and the `large' group, for which $q_i > 1/n$.

The objective is to construct $n$ bins, each with an equal chance ($1/n$) of being chosen. Within each bin, a member from the `small' group (primary outcome) is paired with a portion of a member from the `large' group (alias outcome), such that their combined probabilities sum to $1/n$. To sample from this distribution, we first select a bin randomly. Since each bin has an equal probability of $1/n$, all bins are equally likely to be chosen. Next, we generate another uniform random number to determine whether to select the primary outcome or the alias outcome from the chosen bin.

Mathematically, the setup can be described as follows:
Let $S$ be the set of outcomes with $q_i \leq 1/n$ and $L$ be the set of outcomes with $q_i > 1/n$. The goal of the algorithm is to divide the $q_i$ probabilities such that each bin contains two outcomes. We denote the index of the primary outcome within bin $i$ as $A_i$ and denote the index of the alias outcome as $B_i$. Meanwhile, the probability of choosing the primary outcome in the $i^\mathrm{th}$ bin is $\tilde{q}_i$, and $1 - \tilde{q}_i$  is the probability of choosing the alias outcome.

The filling process can be described with this pseudocode:
\begin{enumerate}
	\item Initiate $\{\tilde{q}_i\} = \{q_i\}$
    \item While $L$ and $S$ are not empty, do the following steps:
    \begin{enumerate}
        \item Randomly pick entry $i$ from small group $S$ and entry $j$ from large group $L$. So, $i$ and $j$ are no longer in the groups.  
        \item Set primary outcome $A_i = i$ and alias outcome $B_i = j$.
        \item Update $\tilde{q}_j = \tilde{q}_j - (1/n - \tilde{q}_i)$.
        \item If $\tilde{q}_j \leq 1/n$, put $j$ into small group $S$.
        \item Otherwise, move $j$ back to large group $L$.
    \end{enumerate}
	\item Now we have $n$ bins filled up to $1/n$. In each bin $i$  the primary outcome $A_i$ occurs with $\tilde{q}_i$ and alias outcome $B_i$ occurs with $1-\tilde{q}_i$.
\end{enumerate}
Step (2.a) iteratively removes entries from groups $S$  and $L$  until both become empty. Since the sum of probabilities must equal one, the presence of an entry in one group guarantees the presence of an entry in the other. In step (2.b), the $i^\mathrm{th}$ bin is filled to a probability of $1/n$  by transferring probability mass from the $j^\mathrm{th}$ bin. Step (2.c) evaluates the remaining probability in the $j^\mathrm{th}$ bin. If it falls below $1/n$, the entry is transferred to group $S$, otherwise, it moves back to the group $L$. This procedure constitutes an `alias table', a data structure containing probabilities for the primary outcomes, $\tilde{q}_i$, indices of the primary outcomes, $A_i$, and indices of the alias outcomes, $B_i$. This setup process is done with a time complexity of $\mathcal{O}(n)$.

The sampling of the alias table can be done with $\mathcal{O}(1)$ using the following pseudocode:
\begin{enumerate}
\item Generate a random integer $i$ in the range $[1, n]$.
\item  Generate a uniform random number $r$ in  interval $[0, 1)$.
\item If $r < \tilde{q}_i$ return the outcome $A_i$, \\ else return the outcome $B_i$.
\end{enumerate}
This method allows us to sample from a probability distribution in $O(1)$ time after an $O(n)$ setup phase \cite{walker1977, vose1991}. This makes it highly efficient for generating numerous samples from the same distribution, as utilized in the next appendix.

\section{Fukui-Todo method \label{app:fukui}}
Using the walker alias technique directly in the clock method is not favorable. To understand why, we explain its direct application in the clock method.  First, we need to set up an alias table for the PDF of the first bound failure event, \textit{i.e.} the alias table for probability distribution $\mathcal{P}(j)$ in Eq. \eqref{eq:distribution}. Sampling this table gives us the location of the first bound failure event, called $j_1$.

If this event is a true rejection, we are finished. Otherwise, we need to find the next bound failure event. This requires setting up an alias table for PDF of the next bound failure event after $j_1$. However, the event $j_1$ could be located in any of the factors. Therefore, in each factor, we would need to set up a distinct alias table for the PDF of the next bound failure event.

Since the size of these alias tables scales as $\mathcal{O}(N)$, storing alias tables for each of the $N$ factors would result in $\mathcal{O}(N^2)$ memory complexity. This can quickly lead to a lot of memory being used, which might not be feasible in real-world applications. In the following, we explain the Fukui-Todo approach which enables the use of the Walker alias method with $\mathcal{O}(N)$ memory requirement \cite{fukui2009}. 

In the clock algorithm, the probability of a failure event at index $j$ is $h_j= 1-p_j= 1-\exp(-\beta[\Delta E_{j}]^+)$.
Here, we set the corresponding bound failure probability to $\hat{h}_j = \max(h_j) = 1 - \exp(-\beta \max(\Delta E_{j}))$.
The idea of the Fukui-Todo approach is to represent this failure probability in terms of a Poisson process. The probability that a Poisson variable takes an integer $k$  is given by
$$
f(k ; \lambda)=\frac{e^{-\lambda} \lambda^k}{k !},
$$
where $\lambda$ is the mean of the distribution. Note that $f(0 ; \lambda)=e^{-\lambda}$ and therefore
$$
\sum_{k=1}^{\infty} f(k ; \lambda)=1-e^{-\lambda}.
$$
which is equal to $\hat{h}_j$, if one puts $\lambda$ to be $\beta \max(\Delta E_{j})$. In other words, if one generates an integer according to the Poisson distribution with mean $\lambda=\beta \max(\Delta E_{j})$, it will take a nonzero value with the probability $\hat{h}_j$.

The essential aspect of representing a failure probability with a Poisson process lies in the fact that the Poisson process is used for random events, and there is no statistical correlation between two separate Poisson processes \cite{fukui2009}. From the perspective of Poisson processes, in each neighbor, there is a process that fails with the mean $\lambda_j$. Since there is no statistical correlation between the processes, all these processes can be considered as a single Poisson process that fails with the total mean  $\lambda_{\text{tot}} = \sum_{j} \lambda_{j}$. This can be shown by the following identity:
$$
\prod_{j=1}^{N-1} f\left(k_{j} ; \lambda_{j}\right)=f\left(k_{\mathrm{tot}} ; \lambda_{\mathrm{tot}}\right) \frac{\left(k_{\mathrm{tot}}\right) !}{k_{1} ! \cdots k_{N-1} !} \prod_{j=1}^{N-1}\left(\frac{\lambda_{j}}{\lambda_{\mathrm{tot}}}\right)^{k_{j}}.
$$
Here, $k_{\text{tot}}=\sum_{j} k_{j}$ is a Poisson random variable  with the mean of  $\lambda_{\text{tot}}$. The left side of the equation represents the probability of having $k_j$ events for each neighbor $j$, independently. The right side, on the other hand, shows the probability for a total of $k_{\text{tot}}$ events across all neighbors, then distributed to each neighbor, weighted proportionally to their individual means, $\lambda_j$. This distribution of events can be efficiently done using the Walker alias method \cite{fukui2009}.

To sample the factorized Metropolis filter in the clock method using the Fukui-Todo algorithm, we first need to generate an alias table for weights $\lambda_j/\lambda_{\text{tot}}$.
Then the sampling can be achieved using the following pseudocode:
\begin{enumerate}
\item Generate a nonnegative integer $k_{\text{tot}}$ according to the Poisson distribution with mean $\lambda_{\text {tot }}$
\item Repeat the following procedure $k_{\text {tot }}$ times:
\begin{enumerate}
\item Choose a neighbor $j$ with the probability ${\lambda_{j}}/{\lambda_{\text {tot}}}$
using the Walker alias method.
\item  If the neighbor has been previously selected, continue to step (2.a).
\item Otherwise, there is a bound failure in this neighbor. Verify if it was a true failure by sampling $h_{j,\text{ver}} = {h_j}/{\hat{h}_j}$ according to Eq. \eqref{eq:hverification}.
\end{enumerate}
\end{enumerate}

Therefore, the Fukui-Todo method only needs a single alias table for the weights $\lambda_j/\lambda_{\text{tot}}$, which can be generated prior to simulation initiation.

\section{Comparison of the dynamic thinning and Fukui-Todo methods \label{app:FTvsDythin}}
\begin{figure}[tbp]
    \centering
	\includegraphics[width=\columnwidth]{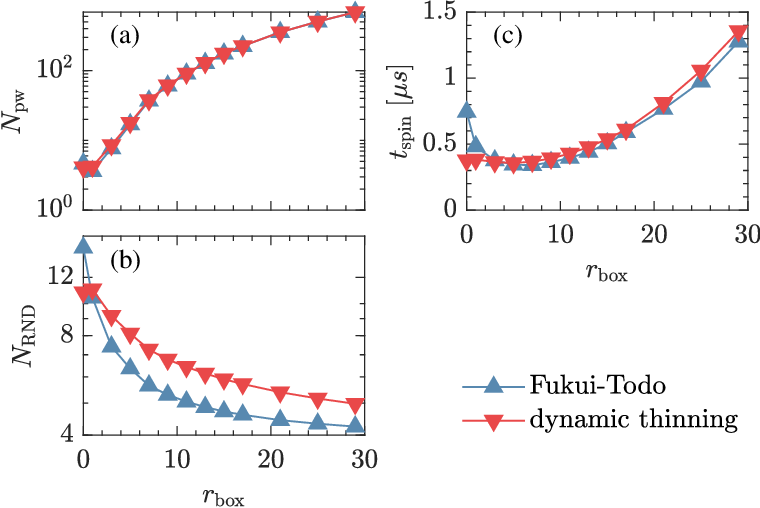}
    \caption{Comparison of the simulation of the clock method equipped with the dynamic thinning and Fukui-Todo techniques in (a) the average number of pairwise interaction evaluations per spin $N_\mathrm{pw}$, (b) the average random number generations per spin $N_\mathrm{RND}$, and (c) the computational time per spin $t_\mathrm{spin}$ \textit{versus} box radius $r_\mathrm{box}$. The simulations are performed at the estimated critical temperature $T_\mathrm{c}=0.849$ for the system size of $N=256 \times 256$. The lines are only a guide to the human eye, and errors are smaller than the size of the symbols.}
    \label{fig:DythinVsFukuiBox}
\end{figure}

In this appendix, we compare the dynamic thinning and Fukui-Todo techniques for identifying failure events in the clock method. We simulated the clock method using these techniques for various box radii at the estimated critical point of the dipolar triangular lattice, $T_\mathrm{c}=0.849$.  Fig.~\ref{fig:DythinVsFukuiBox} presents the results for pairwise energy evaluations and random number generations per spin as a function of box radius. The number of pairwise evaluations ($N_\mathrm{pw}$) shows negligible differences between the techniques.  However, without employing the boxing technique, random number generations per spin ($N_\mathrm{RND}$) is lower for the dynamic thinning method. Notably, $N_\mathrm{RND}$ decreases with increasing box size for both methods. However, the reduction in $N_\mathrm{RND}$ is more pronounced for the Fukui-Todo technique, leading to lower $N_\mathrm{RND}$ for larger boxes.  Consequently, the computational time per spin ($t_\mathrm{spin}$) is lower for dynamic thinning without the boxing technique, while the Fukui-Todo technique becomes more efficient for larger box sizes.

\begin{figure}[tbp]
    \centering
	\includegraphics[width=\linewidth]{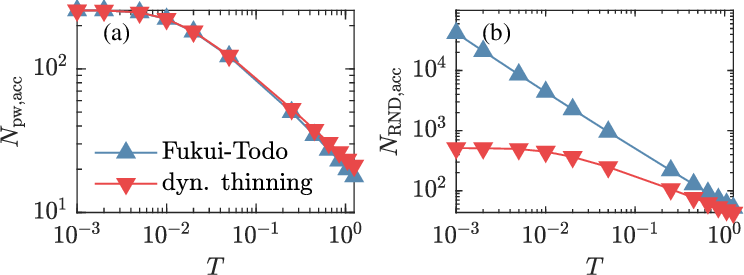}
    \caption{Plots compare the efficiency of the clock method equipped with the dynamic thinning and Fukui-Todo techniques. (a) shows the average pairwise energy evaluations in an accepted move $N_\mathrm{pw,acc}$, and (b) indicates the average random number generations for an accepted move $N_\mathrm{RND,acc}$, in terms of temperature. The system size is $L=16$, and errors are smaller than the data points. The lines serve as a guide to the human eye.}
    \label{fig:DythinVsFukuiTemp}
\end{figure}

For comparison of the dynamic thinning and Fukui-Todo techniques at different temperatures, simulations were conducted with a system length of $L=16$. Fig.~\ref{fig:DythinVsFukuiTemp} depicts the average number of pairwise energy evaluations per accepted move ($N_\mathrm{pw,acc}$) and random number generations per accepted move ($N_\mathrm{RND,acc}$) for both techniques across different temperatures. The results show that $N_\mathrm{pw,acc}$ remains equal for both methods independent of temperature. However, $N_\mathrm{RND,acc}$ for the Fukui-Todo technique exhibits a linear increase in log-log plot ($\sim 1/T$) with decreasing temperature, while it reaches a constant value for dynamic thinning, proportional to the number of particles, $N$. Therefore, both techniques become inefficient at low temperatures.

The findings suggest that the dynamic thinning technique exhibits Higher efficiency at lower temperatures and in simulations with small box sizes. Conversely, the Fukui-Todo technique becomes more effective at higher temperatures and with larger boxes. The results obtained with the SCO method were highly similar to those with the clock method (data not shown).

\section{Reduction of the computational time by boxing near neighbors \label{app:theory}}
In this appendix, we present a theoretical analysis demonstrating how the use of the near-neighbors box can lead to a reduction in computational time. This is achieved by reducing random number generations during the simulation process.  

The average computation time of a spin update $t_\mathrm{spin} = t_\mathrm{MCS} / N$  is estimated by
\begin{equation}
t_\mathrm{spin}  \approx t_\mathrm{pw} {N}_\mathrm{pw} + t_\mathrm{RND} {N}_\mathrm{RND}. \\
\end{equation}
Here $t_\mathrm{pw}$ is the computation time for evaluating a pairwise interaction, and ${N}_\mathrm{pw}$ is the \emph{average} number of pairwise interaction evaluations per spin. $t_\mathrm{RND}$ is the computation time of generating a random number and ${N}_\mathrm{RND}$ represents the \emph{average} random number generations per spin.

Due to the effective field method, the cost of a rejected move is significantly lower than an accepted move. Therefore, $t_\mathrm{spin}$ is primarily dominated by the update time of an accepted move. In an accepted move, the first term arises mainly from updating effective fields for spins in the box, and the second term comes primarily from random numbers generated for the stochastic selection of far spins. So, The average computation time of a spin update can be approximated by 
\begin{equation}
t_\mathrm{spin} \approx P_\mathrm{acc} \left[ t_\mathrm{pw}  N_\mathrm{box} + t_\mathrm{RND} \tilde{N}_\mathrm{far}\right] ,
\end{equation}
where $\tilde{N}_\mathrm{far}$ is the number of stochastically selected far spins and $P_\mathrm{acc}$ is the acceptance ratio.

Next, we consider the box to be a circular region of radius $r_\text{box}$ centered on the changing spin. For the first term, we estimate $N_\text{box}$ by the area of the circle (\textit{i.e.},~$\pi r_\text{box}^2$). In the Fukui-Todo method, the summation of Poisson rates for far interactions, $\lambda_{\text {tot }}=\sum_{j \notin {\text{box}}} \lambda_{j}$ indicates the average of $\tilde{N}_\mathrm{far}$. The values of Poisson rates in free boundaries can be calculated as $\lambda_{j} = \beta \max(\vec{S}_i \cdot \vec{D}_{ij} \cdot \vec{S}_j) = 4\beta	 r_{ij}^{-3}$ where we set $\mu_0/4\pi=1$ and the maximum is taken by diagonalization of dyadic tensor $\vec{D}_{ij}$. Intuitively, the maximum energy difference occurs when the spins $\vec{S}_i$ and $\vec{S}_j$ change from an aligned configuration to become antialigned. In the limit of infinite size, this maximum for free boundaries is identical to that for periodic boundaries. We will then approximate the summation $\lambda_{\text {tot }}=\sum_{j \notin {\text{box}}} 4\beta r_{ij}^{-3}$ by an integral. Using these approximations we have
\begin{align}
t_\mathrm{spin} & \approx t_\mathrm{pw} P_\mathrm{acc} \pi r_\mathrm{box}^2 +  2 t_\mathrm{RND} P_\mathrm{acc}  \int_{r_\mathrm{box}}^{\frac{L}{2}} 4\beta r^{-3} \; (2\pi r)\, \mathrm{d} r \notag \\ 
& \approx  P_\mathrm{acc} \Big[ t_\mathrm{pw}  \pi r_\mathrm{box}^2  +   16 \pi \beta  t_\mathrm{RND} \left( r_\mathrm{box}^{-1} - 2L^{-1} \right) \Big] , \label{eq:t_spin_final}
\end{align} 
where $\beta$ is the inverse temperature, and $L$ is the size of the system. Factor 2 in the second term is due to the use of the Walker alias method which requires two random numbers to sample a distribution.

The Eq. \eqref{eq:t_spin_final} is suitable for a rough estimate of $t_\mathrm{spin}$ as a function of $r_\mathrm{box}$. The equation reveals that minimizing $t_\mathrm{spin}$ leads to a radius, $r_\mathrm{box}$, that minimizes the update time and is independent of the system size. Therefore, if the simulation finds the minimum value of $t_\mathrm{spin}$ with respect to $r_\mathrm{box}$ for a specific system length $L$, this radius is the minimum $t_\mathrm{spin}$ for other sizes as well.

\begin{figure}[tbp]
	\centering
	\includegraphics[width=0.7\columnwidth]{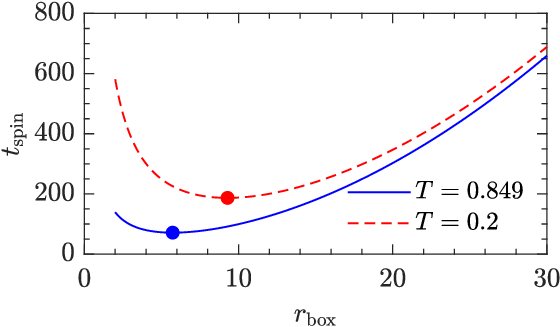}
	\caption{\label{fig:r2plus1r} Theoretical spin update time \textit{vs.} box radius for two different temperatures in the clock method. The typical parameters $L=\infty$ and $t_\text{pw}=1$, $t_\text{RND}=20$, $P_\text{acc}=0.23$ and are used. Filled circles indicate the minima of the plots.} 
\end{figure}

Also, the Eq. \eqref{eq:t_spin_final}  indicates that the minimum for $t_\mathrm{spin}$ in terms of $r_\mathrm{box}$ depends on the temperature. At low temperatures ($\beta \gg 1$ ), the random number generations increase. Consequently, the minimum value of $t_\mathrm{spin}$ occurs for larger values of $r_\mathrm{box}$, as shown in Fig.~\ref{fig:r2plus1r}. Therefore, to minimize $t_\mathrm{spin}$, it is necessary to increase the radius of the box.

Although this relation was derived specifically for the clock method, it can also be applied to the SCO and Tomita methods with minor modifications in the multiplicative factors. The essential characteristic of Eq. \eqref{eq:t_spin_final} is the competition between the $r^2$ and $r^{-1}$ terms. This interplay is crucial because it influences the shape of the $t_\mathrm{spin}$ curves as a function of $r_\mathrm{box}$. However, the precise curves of $t_\mathrm{spin}$ must be determined through numerical simulation.
Therefore, deriving the analytical forms of $t_\mathrm{spin}$ for the SCO and Tomita methods is not necessary in this context.

\section{Stochastic cutoff method \label{app:SCO}} 
In this appendix, we provide a brief explanation of the stochastic cutoff (SCO) method \cite{sasaki2008}, which is based on the stochastic potential switching algorithm \cite{mak2005}.
Suppose we are going to update the spin $i$ of the system. The interaction energy of the spin $i$ with all of its neighbors is given by
\begin{equation}
\mathcal{H}_i = \sum_{j (\ne i)}  V_j(\vec{S}_i, \vec{S}_j),
\end{equation}
where $V_j =  - \vec{S}_i \cdot \vec{D}_{ij} \cdot \vec{S}_j$ is the potential energy between spin $i$ and $j$. In the stochastic cutoff algorithm, $V_j$ is stochastically switches to 0 with probability of $P_j$ as 
\begin{equation}\label{eq:SCOProb}
P_j(\vec{S}_i, \vec{S}_j) = \exp[\beta (V_j(\vec{S}_i, \vec{S}_j) - \ V^{\text{max}}_j)]. \\
\end{equation}
Here $\beta$ is the inverse temperature, $V^{\text{max}}_j$ is a constant equal to (or greater than) the maximum of $V_j(\vec{S}_i, \vec{S}_j)$ over all configurations of $\vec{S}_i$ and $\vec{S}_j$. This switching procedure comes with the price that if the switching fails, the potential will be modified to pseudopotential $\tilde{V}_j$ according to 
\begin{equation}
\tilde{V}_j(\vec{S}_i, \vec{S}_j) = V_j(\vec{S}_i, \vec{S}_j) - \beta^{-1} \log(1 - P_j) \label{eq:SCOPseudoPotent}.
\end{equation}

In order to switch the potentials, a procedure similar to the clock algorithm is followed. We define the probability of switching failure $h_j = 1 - P_j$ and its maximum value as $\hat{h}_j = \max(h_j)$, which is independent of spins configuration. We then switch the potentials using either the dynamic thinning or Fukui-Todo Methods. As the interaction with distant spins is weak, $\hat{h}_j$ for these interactions is small. Consequently, the potentials for many distant spins will switch to zero. However, some of the interactions, mostly from the nearby spins, will fail to switch and will be modified to pseudopotentials.

After switching the potentials, we compute the modified interaction energy of spin $i$ as
\begin{equation}
\mathcal{H}'_i(\vec{S}_i, \vec{S}_j) = \sum_{j \in \text{f.s.}} \tilde{V}_j(\vec{S}_i, \vec{S}_j).
\end{equation}
Here the sum over $j$  only includes spins that are failed to switch (f.s.). Then we propose a change in the spin $i$ to a random direction $\vec{S}'_i$ and compute the change in modified energy using 	
\begin{equation}
\Delta E' = \mathcal{H}'_i(\vec{S}'_i, \vec{S}_j) - \mathcal{H}'_i(\vec{S}_i, \vec{S}_j),
\end{equation}
Finally, we accept this change with the Metropolis filter $P = e^{-\beta [\Delta E']^{+}}$ \cite{sasaki2008}.

In the SCO method, we can use overrelaxation moves without employing the near-neighbors box. In this case the effective field $\vec{H}_i$ is defined as the effective field of the failed-to-switch spins as
\begin{equation}
\vec{H}_i = \sum_{j \in \text{f.s.}} \vec{D}_{ij} \cdot \vec{S}_j.
\end{equation}  
in the other case, with the near-neighbors box, the effective field becomes
\begin{equation}
\vec{H}_i^{\text{box}} = \sum_{j \in \text{box}} \vec{D}_{ij} \cdot \vec{S}_j + \sum_{j \in \text{f.s.}} \vec{D}_{ij} \cdot \vec{S}_j.
\end{equation}  
where the first summation is over all near neighbors in the box, while the second summation is for far neighbors but only includes the failed-to-switch spins.

Theoretically, one can explore the possibility of using advanced boxing technique in the SCO method, similar to the technique employed in the clock method \cite{michel2019}. This can be achieved by writing the interaction energy of spin $i$ with its neighbors as
\begin{equation}
\mathcal{H}_i = \sum_{k=1}^{N_B}  V_k^\mathrm{B}(\vec{S}_i, B_k),
\end{equation}
where the interaction potential with the $k^\mathrm{th}$ box $B_k$ is defined as 
\begin{equation}
V_k^\mathrm{B}(\vec{S}_i, B_k)  = \sum_{j \in B_k}  V_j(\vec{S}_i, \vec{S}_j).
\end{equation}
Then, the $V_j$ in the definition of \eqref{eq:SCOProb} and \eqref{eq:SCOPseudoPotent} will be replaced by $V_k^\mathrm{B}$. 

%\bibliography{revision2,library2,libraryExtra}

\begin{thebibliography}{64}%
\makeatletter
\providecommand \@ifxundefined [1]{%
 \@ifx{#1\undefined}
}%
\providecommand \@ifnum [1]{%
 \ifnum #1\expandafter \@firstoftwo
 \else \expandafter \@secondoftwo
 \fi
}%
\providecommand \@ifx [1]{%
 \ifx #1\expandafter \@firstoftwo
 \else \expandafter \@secondoftwo
 \fi
}%
\providecommand \natexlab [1]{#1}%
\providecommand \enquote  [1]{``#1''}%
\providecommand \bibnamefont  [1]{#1}%
\providecommand \bibfnamefont [1]{#1}%
\providecommand \citenamefont [1]{#1}%
\providecommand \href@noop [0]{\@secondoftwo}%
\providecommand \href [0]{\begingroup \@sanitize@url \@href}%
\providecommand \@href[1]{\@@startlink{#1}\@@href}%
\providecommand \@@href[1]{\endgroup#1\@@endlink}%
\providecommand \@sanitize@url [0]{\catcode `\\12\catcode `\$12\catcode
  `\&12\catcode `\#12\catcode `\^12\catcode `\_12\catcode `\%12\relax}%
\providecommand \@@startlink[1]{}%
\providecommand \@@endlink[0]{}%
\providecommand \url  [0]{\begingroup\@sanitize@url \@url }%
\providecommand \@url [1]{\endgroup\@href {#1}{\urlprefix }}%
\providecommand \urlprefix  [0]{URL }%
\providecommand \Eprint [0]{\href }%
\providecommand \doibase [0]{https://doi.org/}%
\providecommand \selectlanguage [0]{\@gobble}%
\providecommand \bibinfo  [0]{\@secondoftwo}%
\providecommand \bibfield  [0]{\@secondoftwo}%
\providecommand \translation [1]{[#1]}%
\providecommand \BibitemOpen [0]{}%
\providecommand \bibitemStop [0]{}%
\providecommand \bibitemNoStop [0]{.\EOS\space}%
\providecommand \EOS [0]{\spacefactor3000\relax}%
\providecommand \BibitemShut  [1]{\csname bibitem#1\endcsname}%
\let\auto@bib@innerbib\@empty
%</preamble>
\bibitem [{\citenamefont {De'Bell}\ \emph {et~al.}(2000)\citenamefont
  {De'Bell}, \citenamefont {MacIsaac},\ and\ \citenamefont
  {Whitehead}}]{debell2000}%
  \BibitemOpen
  \bibfield  {author} {\bibinfo {author} {\bibfnamefont {K.}~\bibnamefont
  {De'Bell}}, \bibinfo {author} {\bibfnamefont {A.~B.}\ \bibnamefont
  {MacIsaac}},\ and\ \bibinfo {author} {\bibfnamefont {J.~P.}\ \bibnamefont
  {Whitehead}},\ }\bibfield  {title} {\bibinfo {title} {Dipolar effects in
  magnetic thin films and quasi-two-dimensional systems},\ }\href
  {https://doi.org/10.1103/RevModPhys.72.225} {\bibfield  {journal} {\bibinfo
  {journal} {Rev. Mod. Phys.}\ }\textbf {\bibinfo {volume} {72}},\ \bibinfo
  {pages} {225} (\bibinfo {year} {2000})}\BibitemShut {NoStop}%
\bibitem [{\citenamefont {Kechrakos}\ and\ \citenamefont
  {Trohidou}(2008)}]{kechrakos2008}%
  \BibitemOpen
  \bibfield  {author} {\bibinfo {author} {\bibfnamefont {D.}~\bibnamefont
  {Kechrakos}}\ and\ \bibinfo {author} {\bibfnamefont {K.~N.}\ \bibnamefont
  {Trohidou}},\ }\bibfield  {title} {\bibinfo {title} {Dipolar interaction
  effects in the magnetic and magnetotransport properties of ordered
  nanoparticle arrays},\ }\href {https://doi.org/10.1166/jnn.2008.18320}
  {\bibfield  {journal} {\bibinfo  {journal} {J. Nanosci. Nanothechno.}\
  }\textbf {\bibinfo {volume} {8}},\ \bibinfo {pages} {2929} (\bibinfo {year}
  {2008})}\BibitemShut {NoStop}%
\bibitem [{\citenamefont {Karmakar}\ \emph {et~al.}(2011)\citenamefont
  {Karmakar}, \citenamefont {Kumar}, \citenamefont {Rinaldi},\ and\
  \citenamefont {Maruccio}}]{karmakar2011}%
  \BibitemOpen
  \bibfield  {author} {\bibinfo {author} {\bibfnamefont {S.}~\bibnamefont
  {Karmakar}}, \bibinfo {author} {\bibfnamefont {S.}~\bibnamefont {Kumar}},
  \bibinfo {author} {\bibfnamefont {R.}~\bibnamefont {Rinaldi}},\ and\ \bibinfo
  {author} {\bibfnamefont {G.}~\bibnamefont {Maruccio}},\ }\bibfield  {title}
  {\bibinfo {title} {Nano-electronics and spintronics with nanoparticles},\
  }\href {https://doi.org/10.1088/1742-6596/292/1/012002} {\bibfield  {journal}
  {\bibinfo  {journal} {J. Phys.: Conf. Ser.}\ }\textbf {\bibinfo {volume}
  {292}},\ \bibinfo {pages} {012002} (\bibinfo {year} {2011})}\BibitemShut
  {NoStop}%
\bibitem [{\citenamefont {Sun}\ \emph {et~al.}(2004)\citenamefont {Sun},
  \citenamefont {Huang},\ and\ \citenamefont {Nikles}}]{sun2004}%
  \BibitemOpen
  \bibfield  {author} {\bibinfo {author} {\bibfnamefont {X.}~\bibnamefont
  {Sun}}, \bibinfo {author} {\bibfnamefont {Y.}~\bibnamefont {Huang}},\ and\
  \bibinfo {author} {\bibfnamefont {D.~E.}\ \bibnamefont {Nikles}},\ }\bibfield
   {title} {\bibinfo {title} {{FePt} and {CoPt} magnetic nanoparticles film for
  future high density data storage media},\ }\href
  {https://doi.org/10.1504/IJNT.2004.004914} {\bibfield  {journal} {\bibinfo
  {journal} {Int. J. Nanotechnol.}\ }\textbf {\bibinfo {volume} {1}},\ \bibinfo
  {pages} {328} (\bibinfo {year} {2004})}\BibitemShut {NoStop}%
\bibitem [{\citenamefont {Peixoto}\ \emph {et~al.}(2020)\citenamefont
  {Peixoto}, \citenamefont {Magalh{\~a}es}, \citenamefont {Navas},
  \citenamefont {Moraes}, \citenamefont {Redondo}, \citenamefont {Morales},
  \citenamefont {Ara{\'u}jo},\ and\ \citenamefont {Sousa}}]{peixoto2020}%
  \BibitemOpen
  \bibfield  {author} {\bibinfo {author} {\bibfnamefont {L.}~\bibnamefont
  {Peixoto}}, \bibinfo {author} {\bibfnamefont {R.}~\bibnamefont
  {Magalh{\~a}es}}, \bibinfo {author} {\bibfnamefont {D.}~\bibnamefont
  {Navas}}, \bibinfo {author} {\bibfnamefont {S.}~\bibnamefont {Moraes}},
  \bibinfo {author} {\bibfnamefont {C.}~\bibnamefont {Redondo}}, \bibinfo
  {author} {\bibfnamefont {R.}~\bibnamefont {Morales}}, \bibinfo {author}
  {\bibfnamefont {J.~P.}\ \bibnamefont {Ara{\'u}jo}},\ and\ \bibinfo {author}
  {\bibfnamefont {C.~T.}\ \bibnamefont {Sousa}},\ }\bibfield  {title} {\bibinfo
  {title} {Magnetic nanostructures for emerging biomedical applications},\
  }\href {https://doi.org/10.1063/1.5121702} {\bibfield  {journal} {\bibinfo
  {journal} {Appl. Phys. Rev.}\ }\textbf {\bibinfo {volume} {7}},\ \bibinfo
  {pages} {011310} (\bibinfo {year} {2020})}\BibitemShut {NoStop}%
\bibitem [{\citenamefont {Newman}\ \emph {et~al.}(1999)\citenamefont {Newman},
  \citenamefont {Barkema}, \citenamefont {Newman},\ and\ \citenamefont
  {Barkema}}]{newman1999}%
  \BibitemOpen
  \bibfield  {author} {\bibinfo {author} {\bibfnamefont {M.~E.~J.}\
  \bibnamefont {Newman}}, \bibinfo {author} {\bibfnamefont {G.~T.}\
  \bibnamefont {Barkema}}, \bibinfo {author} {\bibfnamefont {M.~E.~J.}\
  \bibnamefont {Newman}},\ and\ \bibinfo {author} {\bibfnamefont {G.~T.}\
  \bibnamefont {Barkema}},\ }\href@noop {} {\emph {\bibinfo {title} {Monte
  {C}arlo Methods in Statistical Physics}}}\ (\bibinfo  {publisher} {Oxford
  University Press},\ \bibinfo {address} {New York},\ \bibinfo {year}
  {1999})\BibitemShut {NoStop}%
\bibitem [{\citenamefont {Kardar}(2007)}]{kardar2007}%
  \BibitemOpen
  \bibfield  {author} {\bibinfo {author} {\bibfnamefont {M.}~\bibnamefont
  {Kardar}},\ }\href {https://doi.org/10.1017/CBO9780511815881} {\emph
  {\bibinfo {title} {Statistical Physics of Fields}}}\ (\bibinfo  {publisher}
  {{Cambridge University Press}},\ \bibinfo {address} {{Cambridge}},\ \bibinfo
  {year} {2007})\BibitemShut {NoStop}%
\bibitem [{\citenamefont {{Flores-Sola}}\ \emph {et~al.}(2017)\citenamefont
  {{Flores-Sola}}, \citenamefont {Weigel}, \citenamefont {Kenna},\ and\
  \citenamefont {Berche}}]{flores-sola2017}%
  \BibitemOpen
  \bibfield  {author} {\bibinfo {author} {\bibfnamefont {E.}~\bibnamefont
  {{Flores-Sola}}}, \bibinfo {author} {\bibfnamefont {M.}~\bibnamefont
  {Weigel}}, \bibinfo {author} {\bibfnamefont {R.}~\bibnamefont {Kenna}},\ and\
  \bibinfo {author} {\bibfnamefont {B.}~\bibnamefont {Berche}},\ }\bibfield
  {title} {\bibinfo {title} {Cluster {M}onte {C}arlo and dynamical scaling for
  long-range interactions},\ }\href
  {https://doi.org/10.1140/epjst/e2016-60338-3} {\bibfield  {journal} {\bibinfo
   {journal} {Eur. Phys. J. Spec. Top.}\ }\textbf {\bibinfo {volume} {226}},\
  \bibinfo {pages} {581} (\bibinfo {year} {2017})}\BibitemShut {NoStop}%
\bibitem [{\citenamefont {Luijten}\ and\ \citenamefont
  {Bl{\"o}te}(1995)}]{luijten1995}%
  \BibitemOpen
  \bibfield  {author} {\bibinfo {author} {\bibfnamefont {E.}~\bibnamefont
  {Luijten}}\ and\ \bibinfo {author} {\bibfnamefont {H.~W.}\ \bibnamefont
  {Bl{\"o}te}},\ }\bibfield  {title} {\bibinfo {title} {Monte {C}arlo method
  for spin models with long-range interactions},\ }\href
  {https://doi.org/10.1142/S0129183195000265} {\bibfield  {journal} {\bibinfo
  {journal} {Int. J. Mod. Phys. C}\ }\textbf {\bibinfo {volume} {06}},\
  \bibinfo {pages} {359} (\bibinfo {year} {1995})}\BibitemShut {NoStop}%
\bibitem [{\citenamefont {Fukui}\ and\ \citenamefont {Todo}(2009)}]{fukui2009}%
  \BibitemOpen
  \bibfield  {author} {\bibinfo {author} {\bibfnamefont {K.}~\bibnamefont
  {Fukui}}\ and\ \bibinfo {author} {\bibfnamefont {S.}~\bibnamefont {Todo}},\
  }\bibfield  {title} {\bibinfo {title} {Order-{N} cluster {M}onte {C}arlo
  method for spin systems with long-range interactions},\ }\href
  {https://doi.org/10.1016/j.jcp.2008.12.022} {\bibfield  {journal} {\bibinfo
  {journal} {J. Comput. Phys.}\ }\textbf {\bibinfo {volume} {228}},\ \bibinfo
  {pages} {2629} (\bibinfo {year} {2009})}\BibitemShut {NoStop}%
\bibitem [{\citenamefont {Janke}(2008)}]{janke2008}%
  \BibitemOpen
  \bibfield  {author} {\bibinfo {author} {\bibfnamefont {W.}~\bibnamefont
  {Janke}},\ }\bibfield  {title} {\bibinfo {title} {Monte {C}arlo methods in
  classical statistical physics},\ }in\ \href
  {https://doi.org/10.1007/978-3-540-74686-7_4} {\emph {\bibinfo {booktitle}
  {Computational {{Many-Particle Physics}}}}},\ \bibinfo {series and number}
  {Lecture {{Notes}} in {{Physics}}},\ \bibinfo {editor} {edited by\ \bibinfo
  {editor} {\bibfnamefont {H.}~\bibnamefont {Fehske}}, \bibinfo {editor}
  {\bibfnamefont {R.}~\bibnamefont {Schneider}},\ and\ \bibinfo {editor}
  {\bibfnamefont {A.}~\bibnamefont {Wei{\ss}e}}}\ (\bibinfo  {publisher}
  {{Springer}},\ \bibinfo {address} {{Berlin}},\ \bibinfo {year} {2008})\ pp.\
  \bibinfo {pages} {79--140}\BibitemShut {NoStop}%
\bibitem [{\citenamefont {Swendsen}\ and\ \citenamefont
  {Wang}(1987)}]{swendsen1987}%
  \BibitemOpen
  \bibfield  {author} {\bibinfo {author} {\bibfnamefont {R.~H.}\ \bibnamefont
  {Swendsen}}\ and\ \bibinfo {author} {\bibfnamefont {J.-S.}\ \bibnamefont
  {Wang}},\ }\bibfield  {title} {\bibinfo {title} {Nonuniversal critical
  dynamics in {M}onte {C}arlo simulations},\ }\href
  {https://doi.org/10.1103/PhysRevLett.58.86} {\bibfield  {journal} {\bibinfo
  {journal} {Phys. Rev. Lett.}\ }\textbf {\bibinfo {volume} {58}},\ \bibinfo
  {pages} {86} (\bibinfo {year} {1987})}\BibitemShut {NoStop}%
\bibitem [{\citenamefont {Wolff}(1989)}]{wolff1989}%
  \BibitemOpen
  \bibfield  {author} {\bibinfo {author} {\bibfnamefont {U.}~\bibnamefont
  {Wolff}},\ }\bibfield  {title} {\bibinfo {title} {Collective {M}onte {C}arlo
  updating for spin systems},\ }\href
  {https://doi.org/10.1103/PhysRevLett.62.361} {\bibfield  {journal} {\bibinfo
  {journal} {Phys. Rev. Lett.}\ }\textbf {\bibinfo {volume} {62}},\ \bibinfo
  {pages} {361} (\bibinfo {year} {1989})}\BibitemShut {NoStop}%
\bibitem [{\citenamefont {Baek}\ \emph {et~al.}(2011)\citenamefont {Baek},
  \citenamefont {Minnhagen},\ and\ \citenamefont {Kim}}]{baek2011}%
  \BibitemOpen
  \bibfield  {author} {\bibinfo {author} {\bibfnamefont {S.~K.}\ \bibnamefont
  {Baek}}, \bibinfo {author} {\bibfnamefont {P.}~\bibnamefont {Minnhagen}},\
  and\ \bibinfo {author} {\bibfnamefont {B.~J.}\ \bibnamefont {Kim}},\
  }\bibfield  {title} {\bibinfo {title} {Kosterlitz-{T}houless transition of
  magnetic dipoles on the two-dimensional plane},\ }\href
  {https://doi.org/10.1103/PhysRevB.83.184409} {\bibfield  {journal} {\bibinfo
  {journal} {Phys. Rev. B}\ }\textbf {\bibinfo {volume} {83}},\ \bibinfo
  {pages} {184409} (\bibinfo {year} {2011})}\BibitemShut {NoStop}%
\bibitem [{\citenamefont {Diaconis}\ \emph {et~al.}(2000)\citenamefont
  {Diaconis}, \citenamefont {Holmes},\ and\ \citenamefont
  {Neal}}]{diaconis2000}%
  \BibitemOpen
  \bibfield  {author} {\bibinfo {author} {\bibfnamefont {P.}~\bibnamefont
  {Diaconis}}, \bibinfo {author} {\bibfnamefont {S.}~\bibnamefont {Holmes}},\
  and\ \bibinfo {author} {\bibfnamefont {R.~M.}\ \bibnamefont {Neal}},\
  }\bibfield  {title} {\bibinfo {title} {Analysis of a nonreversible {M}arkov
  chain sampler},\ }\href@noop {} {\bibfield  {journal} {\bibinfo  {journal}
  {Ann. Appl. Probab.}\ }\textbf {\bibinfo {volume} {10}},\ \bibinfo {pages}
  {726} (\bibinfo {year} {2000})}\BibitemShut {NoStop}%
\bibitem [{\citenamefont {Faizi}\ \emph {et~al.}(2020)\citenamefont {Faizi},
  \citenamefont {Deligiannidis},\ and\ \citenamefont {Rosta}}]{faizi2020}%
  \BibitemOpen
  \bibfield  {author} {\bibinfo {author} {\bibfnamefont {F.}~\bibnamefont
  {Faizi}}, \bibinfo {author} {\bibfnamefont {G.}~\bibnamefont
  {Deligiannidis}},\ and\ \bibinfo {author} {\bibfnamefont {E.}~\bibnamefont
  {Rosta}},\ }\bibfield  {title} {\bibinfo {title} {Efficient irreversible
  {M}onte {C}arlo samplers},\ }\href {https://doi.org/10.1021/acs.jctc.9b01135}
  {\bibfield  {journal} {\bibinfo  {journal} {J. Chem. Theory Comput.}\
  }\textbf {\bibinfo {volume} {16}},\ \bibinfo {pages} {2124} (\bibinfo {year}
  {2020})}\BibitemShut {NoStop}%
\bibitem [{\citenamefont {Turitsyn}\ \emph {et~al.}(2011)\citenamefont
  {Turitsyn}, \citenamefont {Chertkov},\ and\ \citenamefont
  {Vucelja}}]{turitsyn2011}%
  \BibitemOpen
  \bibfield  {author} {\bibinfo {author} {\bibfnamefont {K.~S.}\ \bibnamefont
  {Turitsyn}}, \bibinfo {author} {\bibfnamefont {M.}~\bibnamefont {Chertkov}},\
  and\ \bibinfo {author} {\bibfnamefont {M.}~\bibnamefont {Vucelja}},\
  }\bibfield  {title} {\bibinfo {title} {Irreversible {M}onte {C}arlo
  algorithms for efficient sampling},\ }\href
  {https://doi.org/10.1016/j.physd.2010.10.003} {\bibfield  {journal} {\bibinfo
   {journal} {Physica D}\ }\textbf {\bibinfo {volume} {240}},\ \bibinfo {pages}
  {410} (\bibinfo {year} {2011})}\BibitemShut {NoStop}%
\bibitem [{\citenamefont {Ottobre}(2016)}]{ottobre2016}%
  \BibitemOpen
  \bibfield  {author} {\bibinfo {author} {\bibfnamefont {M.}~\bibnamefont
  {Ottobre}},\ }\bibfield  {title} {\bibinfo {title} {Markov chain {M}onte
  {C}arlo and irreversibility},\ }\href
  {https://doi.org/10.1016/S0034-4877(16)30031-3} {\bibfield  {journal}
  {\bibinfo  {journal} {Rep. Math. Phys.}\ }\textbf {\bibinfo {volume} {77}},\
  \bibinfo {pages} {267} (\bibinfo {year} {2016})}\BibitemShut {NoStop}%
\bibitem [{\citenamefont {Suwa}\ and\ \citenamefont {Todo}(2010)}]{suwa2010}%
  \BibitemOpen
  \bibfield  {author} {\bibinfo {author} {\bibfnamefont {H.}~\bibnamefont
  {Suwa}}\ and\ \bibinfo {author} {\bibfnamefont {S.}~\bibnamefont {Todo}},\
  }\bibfield  {title} {\bibinfo {title} {Markov chain {M}onte {C}arlo method
  without detailed balance},\ }\href
  {https://doi.org/10.1103/PhysRevLett.105.120603} {\bibfield  {journal}
  {\bibinfo  {journal} {Phys. Rev. Lett.}\ }\textbf {\bibinfo {volume} {105}},\
  \bibinfo {pages} {120603} (\bibinfo {year} {2010})}\BibitemShut {NoStop}%
\bibitem [{\citenamefont {Ichiki}\ and\ \citenamefont
  {Ohzeki}(2013)}]{ichiki2013}%
  \BibitemOpen
  \bibfield  {author} {\bibinfo {author} {\bibfnamefont {A.}~\bibnamefont
  {Ichiki}}\ and\ \bibinfo {author} {\bibfnamefont {M.}~\bibnamefont
  {Ohzeki}},\ }\bibfield  {title} {\bibinfo {title} {Violation of detailed
  balance accelerates relaxation},\ }\href
  {https://doi.org/10.1103/PhysRevE.88.020101} {\bibfield  {journal} {\bibinfo
  {journal} {Phys. Rev. E}\ }\textbf {\bibinfo {volume} {88}},\ \bibinfo
  {pages} {020101(R)} (\bibinfo {year} {2013})}\BibitemShut {NoStop}%
\bibitem [{\citenamefont {Ren}\ and\ \citenamefont {Orkoulas}(2006)}]{ren2006}%
  \BibitemOpen
  \bibfield  {author} {\bibinfo {author} {\bibfnamefont {R.}~\bibnamefont
  {Ren}}\ and\ \bibinfo {author} {\bibfnamefont {G.}~\bibnamefont {Orkoulas}},\
  }\bibfield  {title} {\bibinfo {title} {Acceleration of {M}arkov chain {M}onte
  {C}arlo simulations through sequential updating},\ }\href
  {https://doi.org/10.1063/1.2168455} {\bibfield  {journal} {\bibinfo
  {journal} {J. Chem. Phys.}\ }\textbf {\bibinfo {volume} {124}},\ \bibinfo
  {pages} {064109} (\bibinfo {year} {2006})}\BibitemShut {NoStop}%
\bibitem [{\citenamefont {Bernard}\ \emph {et~al.}(2009)\citenamefont
  {Bernard}, \citenamefont {Krauth},\ and\ \citenamefont
  {Wilson}}]{bernard2009}%
  \BibitemOpen
  \bibfield  {author} {\bibinfo {author} {\bibfnamefont {E.~P.}\ \bibnamefont
  {Bernard}}, \bibinfo {author} {\bibfnamefont {W.}~\bibnamefont {Krauth}},\
  and\ \bibinfo {author} {\bibfnamefont {D.~B.}\ \bibnamefont {Wilson}},\
  }\bibfield  {title} {\bibinfo {title} {Event-chain {M}onte {C}arlo algorithms
  for hard-sphere systems},\ }\href
  {https://doi.org/10.1103/PhysRevE.80.056704} {\bibfield  {journal} {\bibinfo
  {journal} {Phys. Rev. E}\ }\textbf {\bibinfo {volume} {80}},\ \bibinfo
  {pages} {056704} (\bibinfo {year} {2009})}\BibitemShut {NoStop}%
\bibitem [{\citenamefont {Michel}\ \emph {et~al.}(2014)\citenamefont {Michel},
  \citenamefont {Kapfer},\ and\ \citenamefont {Krauth}}]{michel2014}%
  \BibitemOpen
  \bibfield  {author} {\bibinfo {author} {\bibfnamefont {M.}~\bibnamefont
  {Michel}}, \bibinfo {author} {\bibfnamefont {S.~C.}\ \bibnamefont {Kapfer}},\
  and\ \bibinfo {author} {\bibfnamefont {W.}~\bibnamefont {Krauth}},\
  }\bibfield  {title} {\bibinfo {title} {Generalized event-chain {M}onte
  {C}arlo: Constructing rejection-free global-balance algorithms from
  infinitesimal steps},\ }\href {https://doi.org/10.1063/1.4863991} {\bibfield
  {journal} {\bibinfo  {journal} {J. Chem. Phys.}\ }\textbf {\bibinfo {volume}
  {140}},\ \bibinfo {pages} {054116} (\bibinfo {year} {2014})}\BibitemShut
  {NoStop}%
\bibitem [{\citenamefont {Michel}\ \emph {et~al.}(2015)\citenamefont {Michel},
  \citenamefont {Mayer},\ and\ \citenamefont {Krauth}}]{michel2015}%
  \BibitemOpen
  \bibfield  {author} {\bibinfo {author} {\bibfnamefont {M.}~\bibnamefont
  {Michel}}, \bibinfo {author} {\bibfnamefont {J.}~\bibnamefont {Mayer}},\ and\
  \bibinfo {author} {\bibfnamefont {W.}~\bibnamefont {Krauth}},\ }\bibfield
  {title} {\bibinfo {title} {Event-chain {M}onte {C}arlo for classical
  continuous spin models},\ }\href
  {https://doi.org/10.1209/0295-5075/112/20003} {\bibfield  {journal} {\bibinfo
   {journal} {Europhys. Lett.}\ }\textbf {\bibinfo {volume} {112}},\ \bibinfo
  {pages} {20003} (\bibinfo {year} {2015})}\BibitemShut {NoStop}%
\bibitem [{\citenamefont {Nishikawa}\ \emph {et~al.}(2015)\citenamefont
  {Nishikawa}, \citenamefont {Michel}, \citenamefont {Krauth},\ and\
  \citenamefont {Hukushima}}]{nishikawa2015}%
  \BibitemOpen
  \bibfield  {author} {\bibinfo {author} {\bibfnamefont {Y.}~\bibnamefont
  {Nishikawa}}, \bibinfo {author} {\bibfnamefont {M.}~\bibnamefont {Michel}},
  \bibinfo {author} {\bibfnamefont {W.}~\bibnamefont {Krauth}},\ and\ \bibinfo
  {author} {\bibfnamefont {K.}~\bibnamefont {Hukushima}},\ }\bibfield  {title}
  {\bibinfo {title} {Event-chain algorithm for the {H}eisenberg model: Evidence
  for $z\simeq1$ dynamic},\ }\href {https://doi.org/10.1103/PhysRevE.92.063306}
  {\bibfield  {journal} {\bibinfo  {journal} {Phys. Rev. E}\ }\textbf {\bibinfo
  {volume} {92}},\ \bibinfo {pages} {063306} (\bibinfo {year}
  {2015})}\BibitemShut {NoStop}%
\bibitem [{\citenamefont {Kapfer}\ and\ \citenamefont
  {Krauth}(2016)}]{kapfer2016}%
  \BibitemOpen
  \bibfield  {author} {\bibinfo {author} {\bibfnamefont {S.~C.}\ \bibnamefont
  {Kapfer}}\ and\ \bibinfo {author} {\bibfnamefont {W.}~\bibnamefont
  {Krauth}},\ }\bibfield  {title} {\bibinfo {title} {Cell-veto {M}onte {C}arlo
  algorithm for long-range systems},\ }\href
  {https://doi.org/10.1103/PhysRevE.94.031302} {\bibfield  {journal} {\bibinfo
  {journal} {Phys. Rev. E}\ }\textbf {\bibinfo {volume} {94}},\ \bibinfo
  {pages} {031302(R)} (\bibinfo {year} {2016})}\BibitemShut {NoStop}%
\bibitem [{\citenamefont {Nishikawa}\ and\ \citenamefont
  {Hukushima}(2016)}]{nishikawa2016}%
  \BibitemOpen
  \bibfield  {author} {\bibinfo {author} {\bibfnamefont {Y.}~\bibnamefont
  {Nishikawa}}\ and\ \bibinfo {author} {\bibfnamefont {K.}~\bibnamefont
  {Hukushima}},\ }\bibfield  {title} {\bibinfo {title} {Phase transitions and
  ordering structures of a model of a chiral helimagnet in three dimensions},\
  }\href {https://doi.org/10.1103/PhysRevB.94.064428} {\bibfield  {journal}
  {\bibinfo  {journal} {Phys. Rev. B}\ }\textbf {\bibinfo {volume} {94}},\
  \bibinfo {pages} {064428} (\bibinfo {year} {2016})}\BibitemShut {NoStop}%
\bibitem [{\citenamefont {Harland}\ \emph {et~al.}(2017)\citenamefont
  {Harland}, \citenamefont {Michel}, \citenamefont {Kampmann},\ and\
  \citenamefont {Kierfeld}}]{harland2017}%
  \BibitemOpen
  \bibfield  {author} {\bibinfo {author} {\bibfnamefont {J.}~\bibnamefont
  {Harland}}, \bibinfo {author} {\bibfnamefont {M.}~\bibnamefont {Michel}},
  \bibinfo {author} {\bibfnamefont {T.~A.}\ \bibnamefont {Kampmann}},\ and\
  \bibinfo {author} {\bibfnamefont {J.}~\bibnamefont {Kierfeld}},\ }\bibfield
  {title} {\bibinfo {title} {Event-chain {M}onte {C}arlo algorithms for three-
  and many-particle interactions},\ }\href
  {https://doi.org/10.1209/0295-5075/117/30001} {\bibfield  {journal} {\bibinfo
   {journal} {Europhys. Lett.}\ }\textbf {\bibinfo {volume} {117}},\ \bibinfo
  {pages} {30001} (\bibinfo {year} {2017})}\BibitemShut {NoStop}%
\bibitem [{\citenamefont {H{\"o}llmer}\ \emph {et~al.}(2022)\citenamefont
  {H{\"o}llmer}, \citenamefont {Maggs},\ and\ \citenamefont
  {Krauth}}]{hollmer2022}%
  \BibitemOpen
  \bibfield  {author} {\bibinfo {author} {\bibfnamefont {P.}~\bibnamefont
  {H{\"o}llmer}}, \bibinfo {author} {\bibfnamefont {A.~C.}\ \bibnamefont
  {Maggs}},\ and\ \bibinfo {author} {\bibfnamefont {W.}~\bibnamefont
  {Krauth}},\ }\bibfield  {title} {\bibinfo {title} {Hard-disk dipoles and
  non-reversible {M}arkov chains},\ }\href {https://doi.org/10.1063/5.0080101}
  {\bibfield  {journal} {\bibinfo  {journal} {J. Chem. Phys.}\ }\textbf
  {\bibinfo {volume} {156}},\ \bibinfo {pages} {084108} (\bibinfo {year}
  {2022})},\ \bibinfo {note} {after submission, we are informed that a new
  article has addressed the limitation of the ECMC method for anisotropic
  interactions: T. Guyon, A. Guillin, and M. Michel, Necessary and sufficient
  symmetries in event-chain Monte Carlo with generalized flows and application
  to hard dimers, J. Chem.Phys. 160, 024117 (2024).}\BibitemShut {Stop}%
\bibitem [{\citenamefont {Michel}\ \emph {et~al.}(2019)\citenamefont {Michel},
  \citenamefont {Tan},\ and\ \citenamefont {Deng}}]{michel2019}%
  \BibitemOpen
  \bibfield  {author} {\bibinfo {author} {\bibfnamefont {M.}~\bibnamefont
  {Michel}}, \bibinfo {author} {\bibfnamefont {X.}~\bibnamefont {Tan}},\ and\
  \bibinfo {author} {\bibfnamefont {Y.}~\bibnamefont {Deng}},\ }\bibfield
  {title} {\bibinfo {title} {Clock {M}onte {C}arlo methods},\ }\href
  {https://doi.org/10.1103/PhysRevE.99.010105} {\bibfield  {journal} {\bibinfo
  {journal} {Phys. Rev. E}\ }\textbf {\bibinfo {volume} {99}},\ \bibinfo
  {pages} {010105(R)} (\bibinfo {year} {2019})}\BibitemShut {NoStop}%
\bibitem [{\citenamefont {Sasaki}\ and\ \citenamefont
  {Matsubara}(2008)}]{sasaki2008}%
  \BibitemOpen
  \bibfield  {author} {\bibinfo {author} {\bibfnamefont {M.}~\bibnamefont
  {Sasaki}}\ and\ \bibinfo {author} {\bibfnamefont {F.}~\bibnamefont
  {Matsubara}},\ }\bibfield  {title} {\bibinfo {title} {Stochastic cutoff
  method for long-range interacting systems},\ }\href
  {https://doi.org/10.1143/JPSJ.77.024004} {\bibfield  {journal} {\bibinfo
  {journal} {J. Phys. Soc. Jpn.}\ }\textbf {\bibinfo {volume} {77}},\ \bibinfo
  {pages} {024004} (\bibinfo {year} {2008})}\BibitemShut {NoStop}%
\bibitem [{\citenamefont {Tomita}(2009{\natexlab{a}})}]{tomita2009}%
  \BibitemOpen
  \bibfield  {author} {\bibinfo {author} {\bibfnamefont {Y.}~\bibnamefont
  {Tomita}},\ }\bibfield  {title} {\bibinfo {title} {Monte {C}arlo study of
  two-dimensional {H}eisenberg dipolar lattices},\ }\href
  {https://doi.org/10.1143/JPSJ.78.114004} {\bibfield  {journal} {\bibinfo
  {journal} {J. Phys. Soc. Jpn.}\ }\textbf {\bibinfo {volume} {78}},\ \bibinfo
  {pages} {114004} (\bibinfo {year} {2009}{\natexlab{a}})}\BibitemShut
  {NoStop}%
\bibitem [{\citenamefont {Komatsu}\ \emph {et~al.}(2018)\citenamefont
  {Komatsu}, \citenamefont {Nonomura},\ and\ \citenamefont
  {Nishino}}]{komatsu2018}%
  \BibitemOpen
  \bibfield  {author} {\bibinfo {author} {\bibfnamefont {H.}~\bibnamefont
  {Komatsu}}, \bibinfo {author} {\bibfnamefont {Y.}~\bibnamefont {Nonomura}},\
  and\ \bibinfo {author} {\bibfnamefont {M.}~\bibnamefont {Nishino}},\
  }\bibfield  {title} {\bibinfo {title} {Temperature-field phase diagram of the
  two-dimensional dipolar {I}sing ferromagnet},\ }\href
  {https://doi.org/10.1103/PhysRevE.98.062126} {\bibfield  {journal} {\bibinfo
  {journal} {Phys. Rev. E}\ }\textbf {\bibinfo {volume} {98}},\ \bibinfo
  {pages} {062126} (\bibinfo {year} {2018})}\BibitemShut {NoStop}%
\bibitem [{\citenamefont {M{\"u}ller}\ \emph {et~al.}(2023)\citenamefont
  {M{\"u}ller}, \citenamefont {Christiansen}, \citenamefont {Schnabel},\ and\
  \citenamefont {Janke}}]{muller2023}%
  \BibitemOpen
  \bibfield  {author} {\bibinfo {author} {\bibfnamefont {F.}~\bibnamefont
  {M{\"u}ller}}, \bibinfo {author} {\bibfnamefont {H.}~\bibnamefont
  {Christiansen}}, \bibinfo {author} {\bibfnamefont {S.}~\bibnamefont
  {Schnabel}},\ and\ \bibinfo {author} {\bibfnamefont {W.}~\bibnamefont
  {Janke}},\ }\bibfield  {title} {\bibinfo {title} {Fast, hierarchical, and
  adaptive algorithm for {M}etropolis {M}onte {C}arlo simulations of long-range
  interacting systems},\ }\href {https://doi.org/10.1103/PhysRevX.13.031006}
  {\bibfield  {journal} {\bibinfo  {journal} {Phys. Rev. X}\ }\textbf {\bibinfo
  {volume} {13}},\ \bibinfo {pages} {031006} (\bibinfo {year}
  {2023})}\BibitemShut {NoStop}%
\bibitem [{\citenamefont {Tomita}(2016)}]{tomita2016}%
  \BibitemOpen
  \bibfield  {author} {\bibinfo {author} {\bibfnamefont {Y.}~\bibnamefont
  {Tomita}},\ }\bibfield  {title} {\bibinfo {title} {Relaxational processes in
  the one-dimensional {I}sing model with long-range interactions},\ }\href
  {https://doi.org/10.1103/PhysRevE.94.062142} {\bibfield  {journal} {\bibinfo
  {journal} {Phys. Rev. E}\ }\textbf {\bibinfo {volume} {94}},\ \bibinfo
  {pages} {062142} (\bibinfo {year} {2016})}\BibitemShut {NoStop}%
\bibitem [{\citenamefont {Girotto}\ \emph {et~al.}(2018)\citenamefont
  {Girotto}, \citenamefont {Malossi}, \citenamefont {{dos Santos}},\ and\
  \citenamefont {Levin}}]{girotto2018}%
  \BibitemOpen
  \bibfield  {author} {\bibinfo {author} {\bibfnamefont {M.}~\bibnamefont
  {Girotto}}, \bibinfo {author} {\bibfnamefont {R.~M.}\ \bibnamefont
  {Malossi}}, \bibinfo {author} {\bibfnamefont {A.~P.}\ \bibnamefont {{dos
  Santos}}},\ and\ \bibinfo {author} {\bibfnamefont {Y.}~\bibnamefont
  {Levin}},\ }\bibfield  {title} {\bibinfo {title} {Lattice model of ionic
  liquid confined by metal electrodes},\ }\href
  {https://doi.org/10.1063/1.5013337} {\bibfield  {journal} {\bibinfo
  {journal} {J. Chem. Phys.}\ }\textbf {\bibinfo {volume} {148}},\ \bibinfo
  {pages} {193829} (\bibinfo {year} {2018})}\BibitemShut {NoStop}%
\bibitem [{\citenamefont {Weis}(2003)}]{weis2003}%
  \BibitemOpen
  \bibfield  {author} {\bibinfo {author} {\bibfnamefont {J.~J.}\ \bibnamefont
  {Weis}},\ }\bibfield  {title} {\bibinfo {title} {Simulation of
  quasi-two-dimensional dipolar systems},\ }\href
  {https://doi.org/10.1088/0953-8984/15/15/311} {\bibfield  {journal} {\bibinfo
   {journal} {J. Phys.: Condens. Matter}\ }\textbf {\bibinfo {volume} {15}},\
  \bibinfo {pages} {S1471} (\bibinfo {year} {2003})}\BibitemShut {NoStop}%
\bibitem [{\citenamefont {Metropolis}\ \emph {et~al.}(2004)\citenamefont
  {Metropolis}, \citenamefont {Rosenbluth}, \citenamefont {Rosenbluth},
  \citenamefont {Teller},\ and\ \citenamefont {Teller}}]{metropolis2004}%
  \BibitemOpen
  \bibfield  {author} {\bibinfo {author} {\bibfnamefont {N.}~\bibnamefont
  {Metropolis}}, \bibinfo {author} {\bibfnamefont {A.~W.}\ \bibnamefont
  {Rosenbluth}}, \bibinfo {author} {\bibfnamefont {M.~N.}\ \bibnamefont
  {Rosenbluth}}, \bibinfo {author} {\bibfnamefont {A.~H.}\ \bibnamefont
  {Teller}},\ and\ \bibinfo {author} {\bibfnamefont {E.}~\bibnamefont
  {Teller}},\ }\bibfield  {title} {\bibinfo {title} {Equation of state
  calculations by fast computing machines},\ }\href
  {https://doi.org/10.1063/1.1699114} {\bibfield  {journal} {\bibinfo
  {journal} {J. Chem. Phys.}\ }\textbf {\bibinfo {volume} {21}},\ \bibinfo
  {pages} {1087} (\bibinfo {year} {2004})}\BibitemShut {NoStop}%
\bibitem [{\citenamefont {Hucht}\ \emph {et~al.}(1995)\citenamefont {Hucht},
  \citenamefont {Moschel},\ and\ \citenamefont {Usadel}}]{hucht1995}%
  \BibitemOpen
  \bibfield  {author} {\bibinfo {author} {\bibfnamefont {A.}~\bibnamefont
  {Hucht}}, \bibinfo {author} {\bibfnamefont {A.}~\bibnamefont {Moschel}},\
  and\ \bibinfo {author} {\bibfnamefont {K.~D.}\ \bibnamefont {Usadel}},\
  }\bibfield  {title} {\bibinfo {title} {{M}onte {C}arlo study of the
  reorientation transition in {H}eisenberg models with dipole interactions},\
  }\href {https://doi.org/10.1016/0304-8853(95)00137-9} {\bibfield  {journal}
  {\bibinfo  {journal} {J. Magn. Magn. Mater.}\ }\textbf {\bibinfo {volume}
  {148}},\ \bibinfo {pages} {32} (\bibinfo {year} {1995})}\BibitemShut
  {NoStop}%
\bibitem [{\citenamefont {Baek}(2011)}]{baek2011a}%
  \BibitemOpen
  \bibfield  {author} {\bibinfo {author} {\bibfnamefont {S.~K.}\ \bibnamefont
  {Baek}},\ }\bibfield  {title} {\bibinfo {title} {Cluster {M}onte {C}arlo
  study of magnetic dipoles},\ }\href {https://doi.org/10.3938/jkps.59.2381}
  {\bibfield  {journal} {\bibinfo  {journal} {J. Korean Phys. Soc.}\ }\textbf
  {\bibinfo {volume} {59}},\ \bibinfo {pages} {2381} (\bibinfo {year}
  {2011})}\BibitemShut {NoStop}%
\bibitem [{\citenamefont {El{\c c}i}\ and\ \citenamefont
  {Weigel}(2013)}]{elci2013}%
  \BibitemOpen
  \bibfield  {author} {\bibinfo {author} {\bibfnamefont {E.~M.}\ \bibnamefont
  {El{\c c}i}}\ and\ \bibinfo {author} {\bibfnamefont {M.}~\bibnamefont
  {Weigel}},\ }\bibfield  {title} {\bibinfo {title} {Efficient simulation of
  the random-cluster model},\ }\href
  {https://doi.org/10.1103/PhysRevE.88.033303} {\bibfield  {journal} {\bibinfo
  {journal} {Phys. Rev. E}\ }\textbf {\bibinfo {volume} {88}},\ \bibinfo
  {pages} {033303} (\bibinfo {year} {2013})}\BibitemShut {NoStop}%
\bibitem [{\citenamefont {Creutz}(1987)}]{creutz1987}%
  \BibitemOpen
  \bibfield  {author} {\bibinfo {author} {\bibfnamefont {M.}~\bibnamefont
  {Creutz}},\ }\bibfield  {title} {\bibinfo {title} {Overrelaxation and {M}onte
  {C}arlo simulation},\ }\href {https://doi.org/10.1103/PhysRevD.36.515}
  {\bibfield  {journal} {\bibinfo  {journal} {Phys. Rev. D}\ }\textbf {\bibinfo
  {volume} {36}},\ \bibinfo {pages} {515} (\bibinfo {year} {1987})}\BibitemShut
  {NoStop}%
\bibitem [{\citenamefont {Peczak}\ and\ \citenamefont
  {Landau}(1993)}]{peczak1993}%
  \BibitemOpen
  \bibfield  {author} {\bibinfo {author} {\bibfnamefont {P.}~\bibnamefont
  {Peczak}}\ and\ \bibinfo {author} {\bibfnamefont {D.~P.}\ \bibnamefont
  {Landau}},\ }\bibfield  {title} {\bibinfo {title} {Dynamical critical
  behavior of the three-dimensional {H}eisenberg model},\ }\href
  {https://doi.org/10.1103/PhysRevB.47.14260} {\bibfield  {journal} {\bibinfo
  {journal} {Phys. Rev. B}\ }\textbf {\bibinfo {volume} {47}},\ \bibinfo
  {pages} {14260} (\bibinfo {year} {1993})}\BibitemShut {NoStop}%
\bibitem [{\citenamefont {Pixley}\ and\ \citenamefont
  {Young}(2008)}]{pixley2008}%
  \BibitemOpen
  \bibfield  {author} {\bibinfo {author} {\bibfnamefont {J.~H.}\ \bibnamefont
  {Pixley}}\ and\ \bibinfo {author} {\bibfnamefont {A.~P.}\ \bibnamefont
  {Young}},\ }\bibfield  {title} {\bibinfo {title} {Large-scale {M}onte {C}arlo
  simulations of the three-dimensional $\mathrm{XY}$ spin glass},\ }\href
  {https://doi.org/10.1103/PhysRevB.78.014419} {\bibfield  {journal} {\bibinfo
  {journal} {Phys. Rev. B}\ }\textbf {\bibinfo {volume} {78}},\ \bibinfo
  {pages} {014419} (\bibinfo {year} {2008})}\BibitemShut {NoStop}%
\bibitem [{\citenamefont {Gvozdikova}\ and\ \citenamefont
  {Zhitomirsky}(2005)}]{gvozdikova2005}%
  \BibitemOpen
  \bibfield  {author} {\bibinfo {author} {\bibfnamefont {M.~V.}\ \bibnamefont
  {Gvozdikova}}\ and\ \bibinfo {author} {\bibfnamefont {M.~E.}\ \bibnamefont
  {Zhitomirsky}},\ }\bibfield  {title} {\bibinfo {title} {A {M}onte {C}arlo
  study of the first-order transition in a {H}eisenberg {FCC}
  antiferromagnet},\ }\href {https://doi.org/10.1134/1.1921323} {\bibfield
  {journal} {\bibinfo  {journal} {J. Exp. Theor. Phys. Lett.}\ }\textbf
  {\bibinfo {volume} {81}},\ \bibinfo {pages} {236} (\bibinfo {year}
  {2005})}\BibitemShut {NoStop}%
\bibitem [{\citenamefont {Stasiak}(2009)}]{stasiak2009}%
  \BibitemOpen
  \bibfield  {author} {\bibinfo {author} {\bibfnamefont {P.}~\bibnamefont
  {Stasiak}},\ }\emph {\bibinfo {title} {Theoretical Studies of Frustrated
  Magnets with Dipolar Interactions}},\ \href
  {https://uwspace.uwaterloo.ca/handle/10012/4877} {\bibinfo {type} {Thesis}},\
  \bibinfo  {school} {University of Waterloo} (\bibinfo {year}
  {2009})\BibitemShut {NoStop}%
\bibitem [{\citenamefont {Ismailzadeh}(2023)}]{onmc}%
  \BibitemOpen
  \bibfield  {author} {\bibinfo {author} {\bibfnamefont {S.}~\bibnamefont
  {Ismailzadeh}},\ }\href {https://github.com/sadeqismailzadeh/ONMC} {\bibinfo
  {title} {Implementation of $\mathcal{O}({N})$ {MC} algorithms}},\ \bibinfo
  {howpublished} {https://github.com/sadeqismailzadeh/ONMC} (\bibinfo {year}
  {2023})\BibitemShut {NoStop}%
\bibitem [{\citenamefont {Fan}\ \emph {et~al.}(2023)\citenamefont {Fan},
  \citenamefont {Zhang},\ and\ \citenamefont {Deng}}]{fan2023}%
  \BibitemOpen
  \bibfield  {author} {\bibinfo {author} {\bibfnamefont {Z.}~\bibnamefont
  {Fan}}, \bibinfo {author} {\bibfnamefont {C.}~\bibnamefont {Zhang}},\ and\
  \bibinfo {author} {\bibfnamefont {Y.}~\bibnamefont {Deng}},\ }\href@noop {}
  {\bibinfo {title} {Clock factorized quantum {M}onte {C}arlo method for
  long-range interacting systems}} (\bibinfo {year} {2023}),\ \Eprint
  {https://arxiv.org/abs/2305.14082} {arxiv:2305.14082 [physics.comp-ph]}
  \BibitemShut {NoStop}%
\bibitem [{\citenamefont {Walker}(1977)}]{walker1977}%
  \BibitemOpen
  \bibfield  {author} {\bibinfo {author} {\bibfnamefont {A.~J.}\ \bibnamefont
  {Walker}},\ }\bibfield  {title} {\bibinfo {title} {An efficient method for
  generating discrete random variables with general distributions},\ }\href
  {https://doi.org/10.1145/355744.355749} {\bibfield  {journal} {\bibinfo
  {journal} {ACM T. Math. Software}\ }\textbf {\bibinfo {volume} {3}},\
  \bibinfo {pages} {253} (\bibinfo {year} {1977})}\BibitemShut {NoStop}%
\bibitem [{\citenamefont {Shanthikumar}(1985)}]{shanthikumar1985}%
  \BibitemOpen
  \bibfield  {author} {\bibinfo {author} {\bibfnamefont {J.~G.}\ \bibnamefont
  {Shanthikumar}},\ }\bibfield  {title} {\bibinfo {title} {Discrete random
  variate generation using uniformization},\ }\href
  {https://doi.org/10.1016/0377-2217(85)90159-6} {\bibfield  {journal}
  {\bibinfo  {journal} {Eur. J. Oper. Res.}\ }\textbf {\bibinfo {volume}
  {21}},\ \bibinfo {pages} {387} (\bibinfo {year} {1985})}\BibitemShut
  {NoStop}%
\bibitem [{\citenamefont {Tomita}(2009{\natexlab{b}})}]{tomita2009a}%
  \BibitemOpen
  \bibfield  {author} {\bibinfo {author} {\bibfnamefont {Y.}~\bibnamefont
  {Tomita}},\ }\bibfield  {title} {\bibinfo {title} {Monte {C}arlo study of
  one-dimensional {I}sing models with long-range interactions},\ }\href
  {https://doi.org/10.1143/JPSJ.78.014002} {\bibfield  {journal} {\bibinfo
  {journal} {J. Phys. Soc. Jpn.}\ }\textbf {\bibinfo {volume} {78}},\ \bibinfo
  {pages} {014002} (\bibinfo {year} {2009}{\natexlab{b}})}\BibitemShut
  {NoStop}%
\bibitem [{\citenamefont {Rastelli}\ \emph {et~al.}(2002)\citenamefont
  {Rastelli}, \citenamefont {Regina},\ and\ \citenamefont
  {Tassi}}]{rastelli2002}%
  \BibitemOpen
  \bibfield  {author} {\bibinfo {author} {\bibfnamefont {E.}~\bibnamefont
  {Rastelli}}, \bibinfo {author} {\bibfnamefont {S.}~\bibnamefont {Regina}},\
  and\ \bibinfo {author} {\bibfnamefont {A.}~\bibnamefont {Tassi}},\ }\bibfield
   {title} {\bibinfo {title} {Planar triangular model with long-range
  interactions},\ }\href {https://doi.org/10.1103/PhysRevB.66.054431}
  {\bibfield  {journal} {\bibinfo  {journal} {Phys. Rev. B}\ }\textbf {\bibinfo
  {volume} {66}},\ \bibinfo {pages} {054431} (\bibinfo {year}
  {2002})}\BibitemShut {NoStop}%
\bibitem [{\citenamefont {Fern{\'a}ndez}\ and\ \citenamefont
  {Alonso}(2007)}]{fernandez2007}%
  \BibitemOpen
  \bibfield  {author} {\bibinfo {author} {\bibfnamefont {J.~F.}\ \bibnamefont
  {Fern{\'a}ndez}}\ and\ \bibinfo {author} {\bibfnamefont {J.~J.}\ \bibnamefont
  {Alonso}},\ }\bibfield  {title} {\bibinfo {title} {Nonuniversal critical
  behavior of magnetic dipoles on a square lattice},\ }\href
  {https://doi.org/10.1103/PhysRevB.76.014403} {\bibfield  {journal} {\bibinfo
  {journal} {Phys. Rev. B}\ }\textbf {\bibinfo {volume} {76}},\ \bibinfo
  {pages} {014403} (\bibinfo {year} {2007})}\BibitemShut {NoStop}%
\bibitem [{\citenamefont {De'Bell}\ \emph {et~al.}(1997)\citenamefont
  {De'Bell}, \citenamefont {MacIsaac}, \citenamefont {Booth},\ and\
  \citenamefont {Whitehead}}]{debell1997}%
  \BibitemOpen
  \bibfield  {author} {\bibinfo {author} {\bibfnamefont {K.}~\bibnamefont
  {De'Bell}}, \bibinfo {author} {\bibfnamefont {A.~B.}\ \bibnamefont
  {MacIsaac}}, \bibinfo {author} {\bibfnamefont {I.~N.}\ \bibnamefont
  {Booth}},\ and\ \bibinfo {author} {\bibfnamefont {J.~P.}\ \bibnamefont
  {Whitehead}},\ }\bibfield  {title} {\bibinfo {title} {Dipolar-induced planar
  anisotropy in ultrathin magnetic films},\ }\href
  {https://doi.org/10.1103/PhysRevB.55.15108} {\bibfield  {journal} {\bibinfo
  {journal} {Phys. Rev. B}\ }\textbf {\bibinfo {volume} {55}},\ \bibinfo
  {pages} {15108} (\bibinfo {year} {1997})}\BibitemShut {NoStop}%
\bibitem [{\citenamefont {Chen}\ \emph {et~al.}(1993)\citenamefont {Chen},
  \citenamefont {Ferrenberg},\ and\ \citenamefont {Landau}}]{chen1993}%
  \BibitemOpen
  \bibfield  {author} {\bibinfo {author} {\bibfnamefont {K.}~\bibnamefont
  {Chen}}, \bibinfo {author} {\bibfnamefont {A.~M.}\ \bibnamefont
  {Ferrenberg}},\ and\ \bibinfo {author} {\bibfnamefont {D.~P.}\ \bibnamefont
  {Landau}},\ }\bibfield  {title} {\bibinfo {title} {Static critical behavior
  of three-dimensional classical {H}eisenberg models: A high-resolution {M}onte
  {C}arlo study},\ }\href {https://doi.org/10.1103/PhysRevB.48.3249} {\bibfield
   {journal} {\bibinfo  {journal} {Phys. Rev. B}\ }\textbf {\bibinfo {volume}
  {48}},\ \bibinfo {pages} {3249} (\bibinfo {year} {1993})}\BibitemShut
  {NoStop}%
\bibitem [{\citenamefont {H{\"o}llmer}\ \emph {et~al.}(2020)\citenamefont
  {H{\"o}llmer}, \citenamefont {Qin}, \citenamefont {Faulkner}, \citenamefont
  {Maggs},\ and\ \citenamefont {Krauth}}]{hollmer2020}%
  \BibitemOpen
  \bibfield  {author} {\bibinfo {author} {\bibfnamefont {P.}~\bibnamefont
  {H{\"o}llmer}}, \bibinfo {author} {\bibfnamefont {L.}~\bibnamefont {Qin}},
  \bibinfo {author} {\bibfnamefont {M.~F.}\ \bibnamefont {Faulkner}}, \bibinfo
  {author} {\bibfnamefont {A.~C.}\ \bibnamefont {Maggs}},\ and\ \bibinfo
  {author} {\bibfnamefont {W.}~\bibnamefont {Krauth}},\ }\bibfield  {title}
  {\bibinfo {title} {{{JeLLyFysh-Version1}}.0~\textemdash{} a {P}ython
  application for all-atom event-chain {M}onte {C}arlo},\ }\href
  {https://doi.org/10.1016/j.cpc.2020.107168} {\bibfield  {journal} {\bibinfo
  {journal} {Comput. Phys. Commun.}\ }\textbf {\bibinfo {volume} {253}},\
  \bibinfo {pages} {107168} (\bibinfo {year} {2020})}\BibitemShut {NoStop}%
\bibitem [{\citenamefont {Binder}\ and\ \citenamefont
  {Heermann}(2010)}]{binder2010}%
  \BibitemOpen
  \bibfield  {author} {\bibinfo {author} {\bibfnamefont {K.}~\bibnamefont
  {Binder}}\ and\ \bibinfo {author} {\bibfnamefont {D.~W.}\ \bibnamefont
  {Heermann}},\ }\href {https://doi.org/10.1007/978-3-642-03163-2} {\emph
  {\bibinfo {title} {{M}onte {C}arlo Simulation in Statistical Physics: An
  Introduction}}},\ \bibinfo {series} {Graduate {{Texts}} in {{Physics}}},
  Vol.~\bibinfo {volume} {0}\ (\bibinfo  {publisher} {{Springer}},\ \bibinfo
  {address} {{Berlin}},\ \bibinfo {year} {2010})\BibitemShut {NoStop}%
\bibitem [{\citenamefont {Binder}(1981)}]{binder1981}%
  \BibitemOpen
  \bibfield  {author} {\bibinfo {author} {\bibfnamefont {K.}~\bibnamefont
  {Binder}},\ }\bibfield  {title} {\bibinfo {title} {Finite size scaling
  analysis of ising model block distribution functions},\ }\href
  {https://doi.org/10.1007/BF01293604} {\bibfield  {journal} {\bibinfo
  {journal} {Z. Phys. B Con. Mat.}\ }\textbf {\bibinfo {volume} {43}},\
  \bibinfo {pages} {119} (\bibinfo {year} {1981})}\BibitemShut {NoStop}%
\bibitem [{\citenamefont {Zheng}(1998)}]{zheng1998}%
  \BibitemOpen
  \bibfield  {author} {\bibinfo {author} {\bibfnamefont {B.}~\bibnamefont
  {Zheng}},\ }\bibfield  {title} {\bibinfo {title} {{M}onte {C}arlo simulations
  of short-time critical dynamics},\ }\href
  {https://doi.org/10.1142/S021797929800288X} {\bibfield  {journal} {\bibinfo
  {journal} {Int. J. Mod. Phys. B}\ }\textbf {\bibinfo {volume} {12}},\
  \bibinfo {pages} {1419} (\bibinfo {year} {1998})}\BibitemShut {NoStop}%
\bibitem [{\citenamefont {Politi}\ \emph {et~al.}(2006)\citenamefont {Politi},
  \citenamefont {Pini},\ and\ \citenamefont {Stamps}}]{politi2006}%
  \BibitemOpen
  \bibfield  {author} {\bibinfo {author} {\bibfnamefont {P.}~\bibnamefont
  {Politi}}, \bibinfo {author} {\bibfnamefont {M.~G.}\ \bibnamefont {Pini}},\
  and\ \bibinfo {author} {\bibfnamefont {R.~L.}\ \bibnamefont {Stamps}},\
  }\bibfield  {title} {\bibinfo {title} {Dipolar ground state of planar spins
  on triangular lattices},\ }\href {https://doi.org/10.1103/PhysRevB.73.020405}
  {\bibfield  {journal} {\bibinfo  {journal} {Phys. Rev. B}\ }\textbf {\bibinfo
  {volume} {73}},\ \bibinfo {pages} {020405(R)} (\bibinfo {year}
  {2006})}\BibitemShut {NoStop}%
\bibitem [{\citenamefont {M{\'o}l}\ and\ \citenamefont
  {Costa}(2014)}]{mol2014}%
  \BibitemOpen
  \bibfield  {author} {\bibinfo {author} {\bibfnamefont {L.~A.~S.}\
  \bibnamefont {M{\'o}l}}\ and\ \bibinfo {author} {\bibfnamefont {B.~V.}\
  \bibnamefont {Costa}},\ }\bibfield  {title} {\bibinfo {title} {The phase
  transition in the anisotropic {H}eisenberg model with long range dipolar
  interactions},\ }\href {https://doi.org/10.1016/j.jmmm.2013.10.023}
  {\bibfield  {journal} {\bibinfo  {journal} {J. Magn. Magn. Mater.}\ }\textbf
  {\bibinfo {volume} {353}},\ \bibinfo {pages} {11} (\bibinfo {year}
  {2014})}\BibitemShut {NoStop}%
\bibitem [{\citenamefont {Krech}\ and\ \citenamefont
  {Luijten}(2000)}]{krech2000}%
  \BibitemOpen
  \bibfield  {author} {\bibinfo {author} {\bibfnamefont {M.}~\bibnamefont
  {Krech}}\ and\ \bibinfo {author} {\bibfnamefont {E.}~\bibnamefont
  {Luijten}},\ }\bibfield  {title} {\bibinfo {title} {Optimized energy
  calculation in lattice systems with long-range interactions},\ }\href
  {https://doi.org/10.1103/PhysRevE.61.2058} {\bibfield  {journal} {\bibinfo
  {journal} {Phys. Rev. E}\ }\textbf {\bibinfo {volume} {61}},\ \bibinfo
  {pages} {2058} (\bibinfo {year} {2000})}\BibitemShut {NoStop}%
\bibitem [{\citenamefont {Vose}(1991)}]{vose1991}%
  \BibitemOpen
  \bibfield  {author} {\bibinfo {author} {\bibfnamefont {M.}~\bibnamefont
  {Vose}},\ }\bibfield  {title} {\bibinfo {title} {A linear algorithm for
  generating random numbers with a given distribution},\ }\href
  {https://doi.org/10.1109/32.92917} {\bibfield  {journal} {\bibinfo  {journal}
  {IEEE T. Software Eng.}\ }\textbf {\bibinfo {volume} {17}},\ \bibinfo {pages}
  {972} (\bibinfo {year} {1991})}\BibitemShut {NoStop}%
\bibitem [{\citenamefont {Mak}(2005)}]{mak2005}%
  \BibitemOpen
  \bibfield  {author} {\bibinfo {author} {\bibfnamefont {C.~H.}\ \bibnamefont
  {Mak}},\ }\bibfield  {title} {\bibinfo {title} {Stochastic potential
  switching algorithm for {M}onte {C}arlo simulations of complex systems},\
  }\href {https://doi.org/10.1063/1.1925273} {\bibfield  {journal} {\bibinfo
  {journal} {J. Chem. Phys.}\ }\textbf {\bibinfo {volume} {122}},\ \bibinfo
  {pages} {214110} (\bibinfo {year} {2005})}\BibitemShut {NoStop}%
\end{thebibliography}


\end{document}